\documentclass[12pt,a4paper]{article}
\usepackage{epsfig,graphicx}
\usepackage{amsmath,amsfonts}
\usepackage{amssymb,cite}
\usepackage{t1enc}

\newcommand{\unit}{\leavevmode\hbox{\small1\kern-3.6pt\normalsize1}}

\def \GeV{{\mathrm{GeV}}}

\newcommand{\pmiss}{p\!\!\!\!\! \not \,\,\,\,}

\begin{document}
\vspace*{-3cm}
\begin{flushright}
FTUAM 03/10\\
FT-UAM/CSIC-03-22\\
hep-ph/0307001 \\
June 2003
\end{flushright}

\begin{center}
\begin{Large}
{\bf Testing the Majorana nature of neutralinos \\[0.2cm]
in supersymmetric theories}
\end{Large}

\vspace{0.5cm}
Juan~A.~Aguilar--Saavedra $^{a}$ and
Ana~M.~Teixeira $^{b}$\\[0.2cm]
{$^{a}$\it Departamento de Física and CFIF, \\
Instituto Superior T\'ecnico, P-1049-001 Lisboa, Portugal} \\
{$^{b}$\it  Departamento de F\'\i sica Te\'orica C-XI and Instituto de
F\'\i sica Te\'orica C-XVI, 
Universidad Aut\'onoma de Madrid, Cantoblanco, E-28049 Madrid, Spain}
\end{center}

\begin{abstract}
We discuss selectron pair production in $e^- e^-$ scattering. These processes
can only occur via $t$-channel neutralino exchange, provided that 
the neutralinos are Majorana fermions. We concentrate on the processes
$e^- e^- \to \tilde e_L \tilde e_L,\tilde e_R \tilde e_R \to 
e^- \tilde \chi_1^0 e^- \tilde \chi_2^0 \to
e^- \tilde \chi_1^0 e^- \tilde \chi_1^0 f \!\bar f$, in which a complete
determination of the final state momenta is possible 
without using the selectron
masses as input. The experimental reconstruction of the selectron masses
in this decay channel provides clear
evidence of the Majorana character of the neutralinos, which is confirmed
by the analysis of the electron energy spectrum.
\end{abstract}



\section{Introduction}
Supersymmetry (SUSY) is one of the most interesting and natural extensions of
the standard model of fundamental interactions (SM). In addition to
providing a solution to many of the SM problems, supersymmetric theories
allow a natural connection to high energy scales, and thus to more fundamental
theories of particle physics~\cite{susy:1, susy:2, susy:3}. 
If SUSY is present at the weak
scale, it is expected that it will be discovered at LHC \cite{LHC}, or even at
the second run of Tevatron \cite{Tevatron}.
After the new particles have been detected, it must be verified that
they are indeed the superpartners of the SM fields. 
It will be crucial to measure their quantum numbers and,
for the case of neutral gauginos, confirm that they are
Majorana fermions. Moreover, masses, mixings, couplings and CP
violating phases must be measured in a model-independent way, so that
the Lagrangian parameters can be determined and the SUSY relations for the
couplings verified.
These tasks are not easy to accomplish, because in general
it will be difficult to separate the different sectors, and
many processes giving the same experimental signatures
will be simultaneously present. In order to disentangle the
different processes and obtain precise measurements,
a high energy $e^+ e^-$ collider like TESLA is
essential to complement the LHC capabilities \cite{TESLA}. In fact, the
TESLA design offers many advantages for SUSY studies, with the
possibility of beam
polarisation and the option of $e^- e^-$, $e
\gamma$ and $\gamma \gamma$ scattering.

\begin{figure}[htb]
\begin{center}
\mbox{\epsfig{file=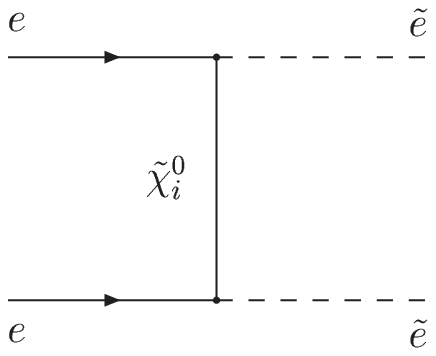,width=4cm,clip=}} \hspace*{15mm}
\mbox{\epsfig{file=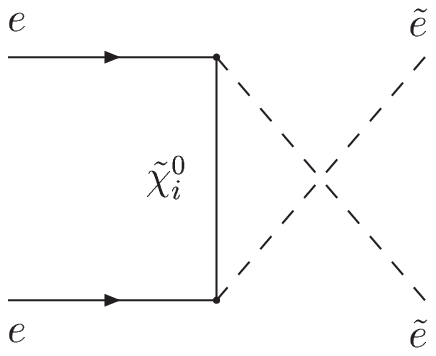,width=4cm,clip=}}
\caption{Feynman diagrams mediating the processes $e^- e^- \to
\tilde e \, \tilde e$.
\label{fig:4legs}}
\end{center}
\end{figure}

Here we focus on the determination of the Majorana nature of
the neutralinos at TESLA. In previous works \cite{petcov,bilenky,gudrid}, it
has been pointed out that the neutralino character can be tested in the
processes $e^+ e^- \to \tilde \chi_1^0 \tilde \chi_i^0 \to \tilde \chi_1^0
\tilde \chi_1^0 \, l^+ l^-$, with $i=2,3,4$ and $l=e,\mu$. These processes give
an experimental signature of two leptons of opposite charge
plus missing energy and momentum\footnote{We assume that the first neutralino
$\tilde \chi_1^0$ is the lightest supersymmetric particle (LSP), 
which is stable if $R$ parity is conserved. Since the $\tilde
\chi_1^0$ is neutral, colourless, and weakly interacting, it 
escapes detection, producing a signature of missing energy and momentum.}, and
the energy distributions of the final state charged leptons are sensitive to
the Dirac or Majorana nature of the decaying neutralino $\tilde \chi_i^0$.
Here we follow a different approach,
studying selectron pair production in $e^- e^-$ collisions
\cite{feng:2001, blochinger:2002}.
In contrast with $e^+ e^-$ scattering, the processes
$e^- e^- \to \tilde e \, \tilde e$ (with $\tilde e = \tilde e_L,\tilde e_R$)
are only mediated by diagrams with $t$-channel neutralino exchange. 
The Majorana nature of the neutralinos is essential for the
nonvanishing of the transition amplitudes,
as can be clearly seen in Fig.~\ref{fig:4legs}.
We note that in some SUSY models it is possible that one or more pairs
of Majorana neutralinos combine to form Dirac (or pseudo-Dirac)
fermions~\cite{sanz1}. This happens when there are two Majorana
neutralinos with the same mass (or approximately the same mass, for
the pseudo-Dirac case) and opposite CP parities. In this situation, the
contributions of these two Majorana neutralinos to selectron pair
production are (almost) equal in modulus with opposite signs, and cancel
(or nearly cancel) in the amplitude.

The dominant decay mode of the selectrons is $\tilde e \to e^- \tilde
\chi_1^0$, leading to a final state with two electrons plus missing
energy and momentum.
The observation of this signal at sizeable rates is a clear indication of
selectron pair production, although in principle it could also
originate from other new physics processes.
In this decay channel, the masses of the selectrons cannot be directly
reconstructed,
due to the presence of two unobservable neutralinos in the final state.
However, the selectron masses can be precisely determined, for instance with a
measurement of the total cross section at energies near threshold
\cite{feng:2001,blochinger:2002}. In the continuum, the study of the
electron energy
spectrum shows that the produced particles are scalars, and the selectron
masses can be extracted from the end points of this distribution \cite{martyn}.
Furthermore, it is possible to construct the minimum kinematically allowed
selectron mass \cite{feng1993}, whose distribution peaks at the
actual selectron masses and hence yields another measurement of these
quantities.
The coincidence of all these measurements provides evidence for selectron
pair production, and disfavours
the interpretation of the $e^- e^-$ signal as originated by other new physics
process.

With the same purpose, in this paper we consider a different decay channel which
offers the advantage that the 4-momenta of {\em all} final state particles can
be determined, and
thus the selectron masses can be directly reconstructed event by event. 
This can be achieved restricting ourselves to $\tilde e_L \tilde e_L$ and 
$\tilde e_R \tilde e_R$
production, and selecting the channel where one of the selectrons decays
to $e^- \tilde \chi_1^0$ and the other selectron decays via
$\tilde e \to e^- \tilde \chi_2^0 \to e^- \tilde \chi_1^0 f \! \bar f$,
with $f \! \bar f$ a $\mu^- \mu^+$ or $q \bar q$ pair.
Although the latter is a relatively rare decay mode, if the selectrons are light
enough to be produced at TESLA, the high luminosity available will allow us to
obtain an observable signal in most cases. 
Using this channel alone, the direct reconstruction of the masses and the
analysis of the two electron energy spectra (one resulting from
$\tilde e \to e^- \tilde \chi_1^0$ and the other 
from $\tilde e \to e^- \tilde \chi_2^0$) provides clear evidence for selectron
pair production. Since the reconstruction method introduced here allows the
determination of the momenta of all final state particles, it can also be used
for further studies of the angular distributions in the production and decay of
the selectrons. In particular, it provides an independent direct verification
that the selectrons are scalar particles. This analysis will be presented
elsewhere \cite{inprep}.

This paper is organised as follows: 
In Section \ref{sec:2} we discuss the
production and subsequent decay of selectron pairs in $e^-e^-$ collisions,
and how these processes are generated.
In Section \ref{sec:3} we briefly describe the model, outlining some important
features, and we select two illustrative scenarios
for $\mu^+ \mu^-$ and $q \bar q$
final states. In Section \ref{sec:4} we describe how the reconstruction
of the selectron masses is achieved and we present our results.
Section \ref{sec:5} is devoted to our conclusions.
In Appendix \ref{sec:A} we collect the mass matrices and 
Lagrangian terms which are relevant for our work. The benefits of beam
polarisation for these processes are displayed in Appendix \ref{sec:B}.

\section{Production and decay of selectron pairs in $\pmb{e^- e^-}$ scattering}
\label{sec:2}

\begin{figure}[htb]
\begin{center}
\begin{tabular}{ccc}
\mbox{\epsfig{file=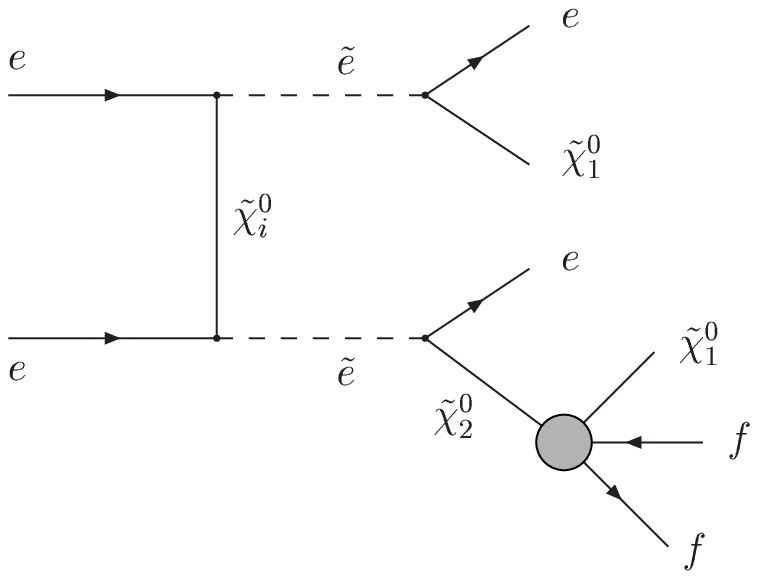,width=6cm,clip=}} & \hspace*{5mm} &
\mbox{\epsfig{file=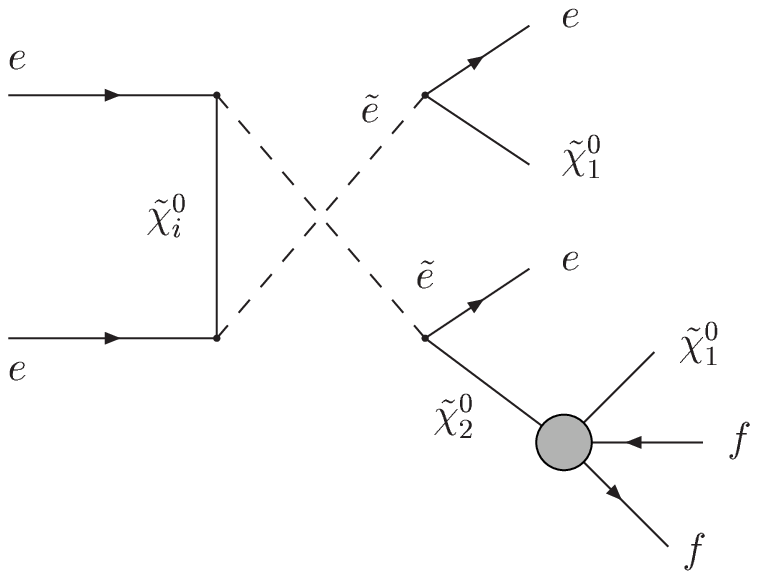,width=6cm,clip=}}
\end{tabular}
\caption{Feynman diagrams for the processes in Eqs.~(\ref{ec:e-e-}).
The 4 neutralinos are exchanged in the $t$ channel, and 
the shaded circles stand for the 7 sub-diagrams mediating the decay of
$\tilde \chi_2^0$, separately shown in Fig.~\ref{fig:X2decay}.}
\label{fig:8legs}
\end{center}
\end{figure}

\begin{figure}[htb]
\begin{center}
\begin{tabular}{ccc}
\mbox{\epsfig{file=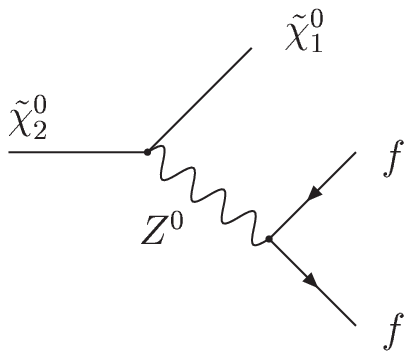,width=3cm,clip=}} & \hspace*{5mm} & 
\mbox{\epsfig{file=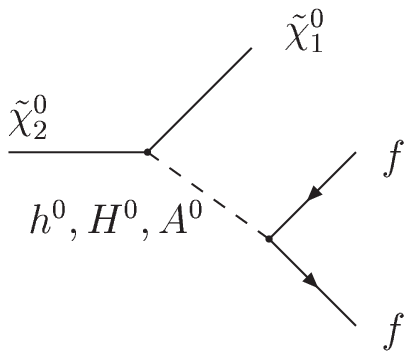,width=3cm,clip=}} \\
(a) & & (b) \\[0.5cm]
\mbox{\epsfig{file=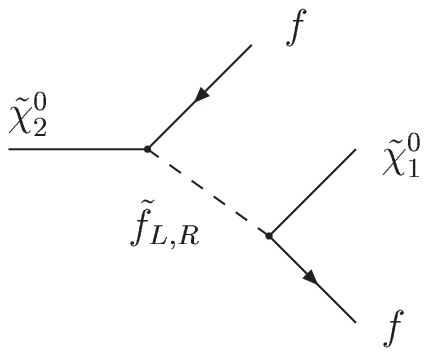,width=3cm,clip=}} & \hspace*{5mm} & 
\mbox{\epsfig{file=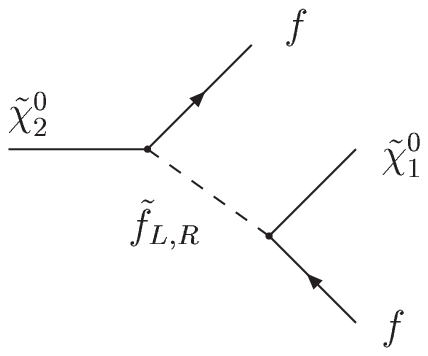,width=3cm,clip=}} \\
(c) & & (d)
\end{tabular}
\caption{Feynman diagrams for the decay of 
$\tilde{\chi}^0_2$, mediated by $Z$ bosons (a), neutral Higgs bosons (b),
and left-and right-handed scalar fermions (c and d).}
\label{fig:X2decay}
\end{center}
\end{figure}

The reconstruction of the selectron masses is only possible for the
production of two identical selectrons,
$e^- e^- \to \tilde e_L \tilde e_L,\tilde e_R \tilde e_R$
(this will be justified in 
the detailed discussion of Section \ref{sec:4}).
For these processes we choose the decay channels
with one selectron decaying via
$\tilde e \to e^- \tilde \chi_1^0$ and the other
$\tilde e \to e^- \tilde \chi_2^0 \to e^- \tilde \chi_1^0 f \!\bar f$.
In addition, we consider mixed selectron production
$e^- e^- \to \tilde e_R \tilde e_L$,
with $\tilde e_R \to e^- \tilde \chi_1^0$,
$\tilde e_L \to e^- \tilde \chi_2^0 \to e^- \tilde \chi_1^0 f \!\bar f$,
since this is the main background to the two processes of
interest\footnote{In the same process $e^- e^- \to \tilde e_R \tilde e_L$, the
channel $\tilde e_L \to e^- \tilde \chi_1^0$,
$\tilde e_R \to e^- \tilde \chi_2^0 \to e^- \tilde \chi_1^0 f \!\bar f$ has a
much smaller cross section because of the smaller branching fractions (see
Section \ref{sec:3}).}.
Each of the processes
\begin{eqnarray}
& & e^- e^- \to \tilde e_L \tilde e_L \to 
e^- \tilde \chi_1^0 e^- \tilde \chi_2^0 \to
e^- \tilde \chi_1^0 e^- \tilde \chi_1^0 f \!\bar f \nonumber \\
& & e^- e^- \to \tilde e_R \tilde e_R \to 
e^- \tilde \chi_1^0 e^- \tilde \chi_2^0 \to
e^- \tilde \chi_1^0 e^- \tilde \chi_1^0 f \!\bar f \nonumber \\
& & e^- e^- \to \tilde e_R \tilde e_L \to 
e^- \tilde \chi_1^0 e^- \tilde \chi_2^0 \to
e^- \tilde \chi_1^0 e^- \tilde \chi_1^0 f \!\bar f
\label{ec:e-e-}
\end{eqnarray}
is mediated by 56 Feynman diagrams, depicted in Figs.
\ref{fig:8legs} and \ref{fig:X2decay}.
Although for clarity the diagrams for the decay of $\tilde{\chi}^0_2$ are
shown
separately, in our computations all the diagrams are summed coherently. We only
consider final states with
$f \!\bar f = \mu^+ \mu^-,q \bar q$, and in the case of $q \bar q$ we sum
$u \bar u$, $d \bar d$, $s \bar s$, $c \bar c$ and $b \bar b$ production,
without flavour tagging. Concerning the remaining channels, in
$\tilde \chi_2^0 \to \tilde \chi_1^0 e^+ e^-$ the multiplicity of
electrons in the final state makes it difficult to identify the electron
resulting from the decay of the $\tilde \chi_2^0$. In
$\tilde \chi_2^0 \to \tilde \chi_1^0 \nu \bar \nu$ the presence of
four undetected particles in the final state yields too many unmeasured momenta
to allow their kinematical determination. The same happens in
$\tilde \chi_2^0 \to \tilde \chi_1^0 \tau^+ \tau^-$, because each of the $\tau$
leptons decays producing one or two neutrinos that escape detection.

Compared with $\tilde e_R \tilde e_L$
production, the other SUSY and SM backgrounds are less important. For
instance, the process $e^- e^- \to \tilde e_R \tilde e_R f \! \bar f$,
with both selectrons decaying to $e^- \tilde \chi_1^0$, has a smaller cross
section, since it has the same number of vertices as $\tilde e_R \tilde e_L$
(both amplitudes are proportional to $g^6$) and does not have the enhancement
of the $\tilde \chi_2^0$ near its mass shell.
The main SM background is $e^- e^- \to e^- e^- \nu \bar \nu f \! \bar f$,
including resonant $e^- e^- \to W^- e^- \nu f \! \bar f$
and $e^- e^- \to e^- e^- Z f \! \bar f$ production.
This process has a small cross section and is separable from the
$\tilde e_L \tilde e_L$, $\tilde e_R \tilde e_R$ signals, which have
larger values for the missing energy and momentum.
One might also consider the process
$e^- e^- \to W^- W^- f \! \bar f$ mediated by
heavy Majorana neutrinos. The cross section for the
latter is very constrained by present limits on
neutrinoless double beta decay \cite{gluza}, and
this process could also be separated from selectron pair production,
using the missing energy and momentum.

We calculate the full matrix elements for the resonant processes in
Eqs.~(\ref{ec:e-e-}) with {\tt HELAS} \cite{helas}, using the Feynman rules for
Majorana fermions given in Ref.~\cite{denner}. 
The inclusion of next-to-leading order
corrections is not necessary in this work, since it only increases
slightly the cross sections \cite{freitas}. We assume a centre of mass (CM)
energy $E_\mathrm{CM} = 2E = 500$ GeV for the $e^- e^-$ collisions, and an
integrated luminosity of 100 fb$^{-1}$.
In order to take into account the
effect of initial state radiation (ISR) on the effective beam energies, we
convolute the differential cross section evaluated for fractions $x_1$, $x_2$
of the beam energies with ``structure functions'' $D_\mathrm{ISR}(x_1)$ and
$D_\mathrm{ISR}(x_2)$, namely
\begin{equation}
d \sigma =  \int_0^1 d\sigma(x_1 E,x_2 E) D_\mathrm{ISR}(x_1)
D_\mathrm{ISR}(x_2) dx_1 dx_2 \;.
\end{equation}
The explicit expression for $D_\mathrm{ISR}$ used is \cite{isr}
\begin{eqnarray}
D_\mathrm{ISR}(x) & = & \frac{\eta}{2} (1-x)^{\frac{\eta}{2}-1}
\frac{e^{\frac{\eta}{2} \left( \frac{3}{4}-\gamma
\right)}}{\Gamma \left( 1 + \frac{\eta}{2} \right)}
\nonumber \\
& & \times \left[ \frac{1}{2} ( 1+x^2 ) - \frac{\eta}{8}
\left( \frac{1}{2} (1+3x^2) \log x - (1-x)^2 \right) \right] \;,
\end{eqnarray}
where
\begin{equation}
\eta(s) = -6 \log \left[ 1- \frac{\alpha_0}{3 \pi} \log \frac{s}{m_e^2} \right]
\;,
\end{equation}
$\gamma$ is the Euler constant, $\alpha_0=1/137$ the fine structure
constant, $s=E_\mathrm{CM}^2$ and $m_e$ the electron mass.
The effect of beamstrahlung is taken into
account in a similar way using the ``structure function'' \cite{peskin}
\begin{equation}
D_\mathrm{BS}(x) = e^{-N} \left[ \delta(x-1) +
\frac{e^{-\kappa(1-x)/x}}{x(1-x)} h(y) \right] \;,
\end{equation}
where \cite{BS2}
\begin{equation}
h(y) = \sum_{n=1}^{\infty} \frac{y^n}{n! \, \Gamma(n/3)} \;,
\end{equation}
with $y=N[\kappa \, (1-x)/x]^{1/3}$. For large $y$, $h(y)$ has the asymptotic
expansion
\begin{equation}
h(y) = \left( \frac{3z}{8\pi} \right)^\frac{1}{2} e^{4z}
\left[ 1- \frac{35}{288 \, z} - \frac{1295}{16588 \, z^2} + \cdots \right] \;,
\end{equation}
with $z=(y/3)^{3/4}$. We take $N = 0.555$, $\kappa = 0.0693$ \cite{peskin}
for our evaluations.

In order to simulate the calorimeter and tracking resolution, we perform a
Gaussian smearing
of the energies of electrons ($e$), jets ($j$) and muons ($\mu$) using the 
specifications in the TESLA Technical Design Report \cite{tesla2}
\begin{equation}
\frac{\Delta E^e}{E^e} = \frac{10\%}{\sqrt{E^e}} \oplus 1 \% \;, \quad
\frac{\Delta E^j}{E^j} = \frac{50\%}{\sqrt{E^j}} \oplus 4 \% \;, \quad
\frac{\Delta E^\mu}{E^\mu} = 0.02 \% \, E^\mu \;,
\end{equation}
where the energies are in GeV and the two terms are added in quadrature. We
apply ``detector'' cuts on transverse momenta, $p_T \geq 10$ GeV, and
pseudorapidities $|\eta| \leq 2.5$, the latter corresponding to polar angles
$10^\circ \leq \theta \leq 170^\circ$. We also reject events in which the
leptons and/or jets are not isolated, requiring that the distances in
$(\eta,\theta)$ space $\Delta R = \sqrt{\Delta \eta^2+\Delta \theta^2}$
satisfy $\Delta R \geq 0.4$.
We do not require specific trigger conditions, and we assume that the
presence of charged leptons with high transverse momentum will suffice.
Finally, for the Monte Carlo integration in 8-body phase space we use
{\tt RAMBO} \cite{rambo}.

\section{Overview of the model and selection of scenarios}
\label{sec:3}

The aim of the present work is to demostrate that the selectron masses can be
successfully reconstructed with the method here proposed. It is nevertheless
important to show that this can be done in realistic scenarios.
In this section, we outline the main features of the model, and 
investigate the behaviour of the selectron production cross sections and decay
rates as a function of the free parameters of the model.

In our analysis we consider the minimal supersymmetric standard model (MSSM),
where a minimal number of fields is introduced and $R$-parity is conserved. 
For simplicity, 
we assume an underlying minimal supergravity (mSUGRA)~\cite{msugra} framework.
At the grand unification scale, $M_{GUT}\simeq 2 \times 10^{16}$ GeV, 
relations of universality among
the several soft breaking terms allow the model to be described by
five independent parameters: the common soft scalar mass ($m_0$), the
unified gaugino mass ($m_{1/2}$), the universal trilinear coupling ($A_0$), 
the ratio of the two neutral
Higgs vacuum expectation values ($\tan \beta= v_2/v_1$) and the 
sign of the bilinear Higgs term, $\operatorname{sign} (\mu)$. 
In our work we do not address supersymmetric CP
violation, thus we assume that $m_{1/2}$ and $A_0$ are real.
The low-energy Lagrangian is derived by means of the
renormalisation group equations, which are numerically solved to
the two-loop order using {\tt SPheno}~\cite{spheno}. In addition, 
{\tt SPheno} is used to obtain the masses and mixings of SUSY
particles, as well as some decay widths. The mass matrices for sfermions
and neutralinos and the required interaction terms
are presented in Appendix~\ref{sec:A}.

It is important to clarify here some issues, which are 
relevant for the subsequent discussion of selectron production and
decay.
The first concerns flavour and chirality mixing in the
slepton sector. For the case of the first two generations, the 
$LR$ term in the slepton mass matrix (Eq.~(\ref{slepton:mass:2ELR}))
is significantly suppressed by the electron and muon Yukawa couplings. 
The smallness of $LR$ mixing
for selectrons and smuons is fairly insensitive to
the particular value of $A_0$, and  
for simplicity we choose $A_0=0$ in our numerical analysis.
This typically translates into
$\theta_{\tilde{e}} \sim \mathcal{O}(10^{-4})$ and 
$\theta_{\tilde{\mu}} \sim \mathcal{O}(10^{-2})$ 
(cf. Eq.~(\ref{mass:slepton:theta})), so that 
$\tilde e_L \tilde e_R$ and $\tilde \mu_L \tilde \mu_R$ mixing 
can indeed be neglected. Therefore, for 
the first two generations the mass eigenstates are the chiral 
$(\tilde{e},\tilde{\mu})_L$, $(\tilde{e},\tilde{\mu})_R$ states. 
Moreover, leptonic interactions are assumed to be flavour-conserving, so only
selectrons are produced in $e^- e^-$ collisions.

As afore mentioned, we assume that the lightest neutralino 
is the LSP. Depending on the underlying model, the 
$\tilde{\chi}^0_1$ can be gaugino- or higgsino-like, or a balanced
admixture of the latter interaction eigenstates. 
Here we shall only consider cases where $\tilde{\chi}^0_1$ is 
gaugino-like and, more specifically, dominated by the bino component,
a situation common to a vast region of the mSUGRA parameter space. 
In our case, the second neutralino, $\tilde{\chi}^0_2$, is dominated
by the wino component, and thus its coupling to the right-handed
selectron $\tilde e_R$ is very small. This fact has a significant impact in
our analysis, since in this situation the $\tilde e_R$ mainly decays to  
$e^- \, \tilde{\chi}^0_1$, even when the $e^- \, \tilde{\chi}^0_2$ 
channel is kinematically allowed, so that the $\tilde e_R \tilde e_R$ process
in Eq.~(\ref{ec:e-e-}) has a very small cross section.
We emphasise that the reconstruction method proposed in
this paper is not sensitive to the 
nature of the lightest neutralinos. However, the cross sections of the
processes studied will be significantly affected, given that the relevant  
$\tilde{\chi}^0_1$-fermion-sfermion couplings are suppressed
if the lightest neutralinos are higgsino-like.

The numerical studies presented in this section do not aim to be a complete scan
of the available parameter space, but rather an illustrative analysis of the
impact of the variation of the mSUGRA parameters on the selectron production and
decay cross sections.
In the following, we will simultaneously try to satisfy the 
direct and indirect bounds on the masses of the particles \cite{PDB}
and still obtain an adequate spectrum for sleptons and
neutralinos, so that the processes in Eq.~(\ref{ec:e-e-}) can be observed
at TESLA with a CM energy of 500 GeV.

\begin{figure}[htb]
\begin{center}
\epsfig{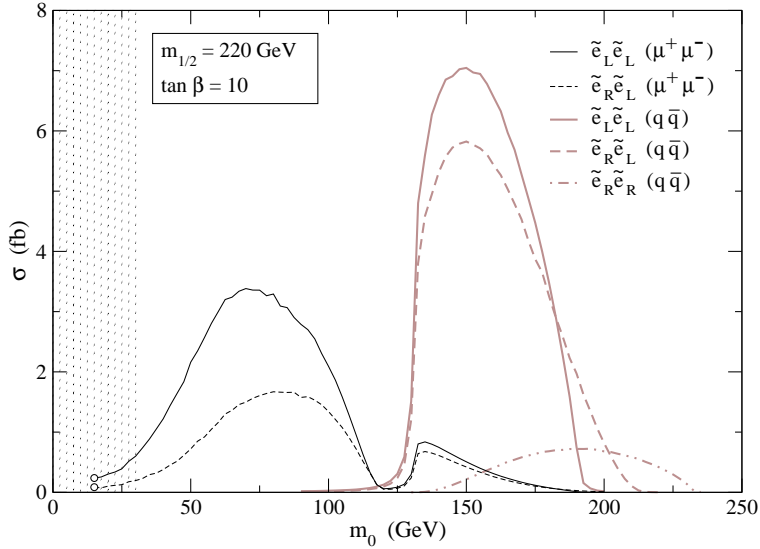}
\caption{Cross sections (in fb) for 
$e^- e^- \to \tilde e \tilde e \to 
e^- \tilde \chi_1^0 e^- \tilde \chi_2^0
\to e^- \tilde \chi_1^0  e^- \tilde \chi_1^0 f \!\bar f$
(with unpolarised beams)
as a function of $m_0$, for $m_{1/2}=220$ GeV and $\tan \beta=10$.}
\label{fig:sigma:m0}
\end{center}
\end{figure}

We begin the selection of the scenarios by 
considering the dependence of the studied signals on $m_0$. 
This is shown in Fig.~\ref{fig:sigma:m0}, 
for $\tilde e_L \tilde e_L$, $\tilde e_R \tilde e_L$ and 
$\tilde e_R \tilde e_R$ production 
(full, dashed and dot-dashed lines) and 
for the $\mu^+ \mu^-$ and $q \bar{q}$ 
channels (black and grey lines, respectively).
We have taken $m_{1/2}=220\ \GeV$, $\tan \beta=10$, and $\mu >0$.
The shaded area on the left corresponds to
values of $m_0$ excluded by the current experimental bounds
on $m_{\tilde{e}_R}$~\cite{PDB}. In addition, the end points of the curves
marked with a circle correspond to the values of $m_0$ below which the lightest
neutralino is not the LSP. In this section we consider unpolarised beams; the
effect of longitudinal polarisation is discussed in Appendix~\ref{sec:B}.

From this plot we notice that only for $m_0 \geq 140$ GeV 
the $\tilde e_R$ is heavy
enough to decay into the second neutralino.
The shape of the curves reflects the strong dependence on $m_0$ of
the production cross sections and branching ratios for
$\tilde e_L \to \tilde e^- \chi_2^0$ and
$\tilde \chi_2^0 \to \tilde \chi_1^0 f \! \bar f$.
In the latter case, there are several competing Feynman
diagrams \cite{bartl,djouadi}, mediated by the $Z$ boson and sfermions, that
contribute to the decay
of the $\tilde \chi_2^0$ (the contribution of the diagrams with
a Higgs boson is small for this value of $\tan \beta$). Three different
regimes can be distinguished:
\begin{itemize}
\item For $m_0 \lesssim 120$ GeV, the second neutralino decays predominantly to
charged leptons, $\tilde \chi_2^0 \to \tilde \chi_1^0 l^+ l^-$.
The decay amplitudes are dominated by diagrams (c) and (d)
in Fig.~\ref{fig:X2decay}, with the exchange of on-shell right-handed 
sleptons, in particular the decay $\tilde \chi_2^0 \to \tilde
\chi_1^0 \mu^+ \mu^-$ is
dominated by $\tilde \mu_R$ exchange. The contribution of 
diagram (a) with an on-shell
$Z$ boson is less important, due to the smallness of the 
$Z \tilde \chi_2^0 \tilde \chi_1^0$ coupling. 
In this region of the parameter space,
the decays to $\tilde \chi_1^0 q \bar q$ are very suppressed, 
not only because of the small coupling of the $Z$ boson
to $\tilde \chi_1^0$ and $\tilde \chi_2^0$, but also due to
the heavy squark masses, $m_{\tilde q} \gtrsim 400$ GeV. 
\item For $m_0 \sim 130$ GeV, the $\tilde e_R$ and $\tilde \mu_R$ are
heavier than the $\tilde \chi_2^0$, while the $\tilde \tau_R$ is
lighter. Then, the $\tilde \chi_2^0$ decays almost exclusively to
$\tilde \chi_1^0 \tau^+ \tau^-$, mediated by an on-shell $\tilde
\tau_R$,  diagrams (c) and (d) in Fig.~\ref{fig:X2decay}. 
(As remarked in Section~\ref{sec:2}, 
in the $\tau^+ \tau^-$ channel the selectron masses 
cannot be reconstructed, and thus we do not consider these final
states in the plots presented throughout this section.)
\item For $m_0 \gtrsim 140$ GeV, the right-handed sleptons
(including $\tilde \tau_R$) are heavier than $\tilde \chi_2^0$, and the
$Z$-exchange diagram in Fig.~\ref{fig:X2decay} dominates, yielding a large
branching ratio for $\tilde \chi_2^0 \to \tilde \chi_1^0 q \bar q$.
\end{itemize}

For $m_0 \gtrsim 200$ GeV, the production of $\tilde e_L \tilde e_L$
and $\tilde e_L \tilde e_R$ is not possible with a CM energy of
500 GeV, and only $\tilde e_R \tilde e_R$ pairs are produced.
From inspection of Fig.~\ref{fig:sigma:m0} we select two values
$m_0 = 80$ GeV and $m_0 = 160$ GeV, approximately in the centre
of the first and third regions, that will serve to illustrate 
the reconstruction of the selectron
masses for $\mu^+ \mu^-$ and $q \bar q$ final states. The sets of
parameters for these two mSUGRA scenarios are summarised in Table \ref{tab:in}.
Both of them are in agreement with the bounds 
arising from the $b \to s \gamma$ decay~\cite{PDB}, and they are in
the region of the $(m_{1/2},m_0)$ plane favoured by $(g-2)_\mu$  at the
$2 \, \sigma$ level \cite{ellis}. The resulting selectron and neutralino masses
and widths, as well as the relevant branching ratios,
are collected in Table \ref{tab:out} for each of these scenarios.
In scenario 1 the $\tilde e_R$ is lighter than the $\tilde \chi_2^0$, and
the decay $\tilde e_R \to e^- \tilde \chi_2^0$ is not possible. Hence, in this
scenario we only consider the $\tilde e_L \tilde e_L$ signal.

\begin{table}[htb]
\begin{center}
\begin{tabular}{ccccc}
Parameter & ~ & Scenario 1 & ~ & Scenario 2 \\
\hline
$m_{1/2}$ & & 220 & & 220 \\
$m_0$ & & 80 & & 160 \\
$A_0$ & & 0 & & 0 \\
$\tan \beta$ & & 10 & & 10 \\
$\mathrm{sign} \, \mu$ & & $+$ & & $+$ 
\end{tabular}
\caption{Input parameters for the two mSUGRA scenarios to be 
considered in Section
\ref{sec:4}. The values of $m_{1/2}$, $m_0$ and $A_0$ are in GeV.
\label{tab:in}}
\end{center}
\end{table}

\begin{table}[htb]
\begin{center}
\begin{tabular}{cccc}
 & Scenario 1 & ~ & Scenario 2 \\
\hline
$m_{\tilde e_L}$ & 181.0 & & 227.4 \\
$\Gamma_{\tilde e_L}$ & 0.25 & & 0.85 \\
$m_{\tilde e_R}$ & 123.0 & & 185.0 \\
$\Gamma_{\tilde e_R}$ & 0.17 & & 0.58 \\
$m_{\tilde \chi_1^0}$ & 84.0 & & 84.3 \\
$m_{\tilde \chi_2^0}$ & 155.8 & & 156.4 \\
$\Gamma_{\tilde \chi_2^0}$ & 0.023 & & $1.50 \times 10^{-5}$ \\
$m_{\tilde \chi_3^0}$ & 309.4 & & 310.0 \\
$\Gamma_{\tilde \chi_3^0}$ & 1.48 & & 1.43 \\
$m_{\tilde \chi_4^0}$ & 330.4 & & 331.1 \\
$\Gamma_{\tilde \chi_4^0}$ & 2.30 & & 2.01 \\
$\mathrm{Br}(\tilde e_L \to e^- \tilde \chi_1^0)$ & 41.8 \% & & 18.3 \% \\
$\mathrm{Br}(\tilde e_L \to e^- \tilde \chi_2^0)$ & 20.6 \% & & 30.8 \% \\
$\mathrm{Br}(\tilde e_R \to e^- \tilde \chi_1^0)$ & 100 \% & & 99.7 \% \\
$\mathrm{Br}(\tilde e_R \to e^- \tilde \chi_2^0)$ & $\simeq 0$ & & 0.3 \% \\
$\mathrm{Br}(\tilde \chi_2^0 \to \tilde \chi_1^0 \mu^+ \mu^-)$ & 10.3 \% & &
 3.9 \% \\
$\mathrm{Br}(\tilde \chi_2^0 \to \tilde \chi_1^0 q \bar q)$ & $\simeq 0$ & &
 69.2 \% \\
\end{tabular}
\caption{Some relevant quantities in the two mSUGRA scenarios
considered in Section \ref{sec:4}. The masses and widths are in GeV.
\label{tab:out}}
\end{center}
\end{table}

\begin{figure}[htb]
\begin{center}
\mbox{\epsfig{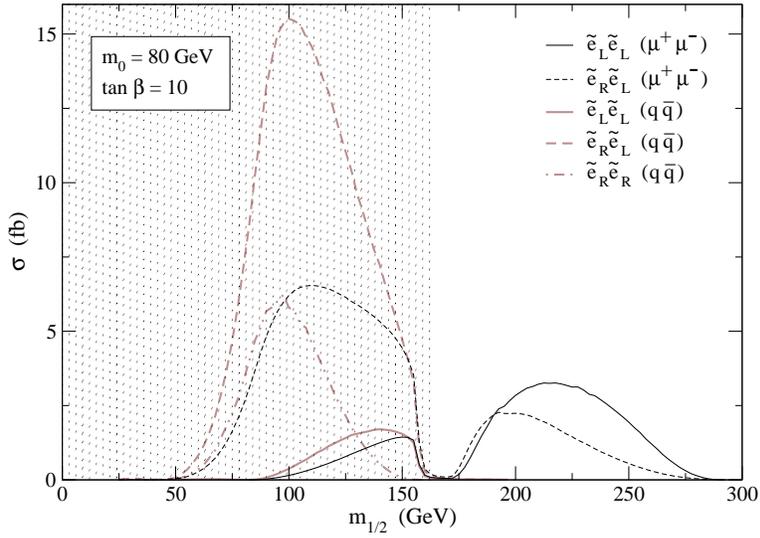}}
\caption{Cross sections (in fb) for 
$e^- e^- \to \tilde e \tilde e \to 
e^- \tilde \chi_1^0 e^- \tilde \chi_2^0
\to e^- \tilde \chi_1^0  e^- \tilde \chi_1^0 f \!\bar f$
(with unpolarised beams)
as a function of $m_{1/2}$, for $m_0 = 80$ GeV and $\tan \beta = 10$.}
\label{fig:sigma:m12}
\end{center}
\end{figure}

For completeness, we investigate the dependence of the cross sections 
on the remaining parameters.
For scenario 1, and leaving all other parameters as in Table~\ref{tab:in},
we plot in Fig.~\ref{fig:sigma:m12} the cross sections for different values of
$m_{1/2}$. The shaded area on the left
corresponds to values of $m_{1/2}$
that yield a $\tilde \chi_1^0$ mass excluded by present limits~\cite{PDB}. 
Note that the depression occurring around $m_{1/2} = 160$
GeV has the same origin as the one observed in Fig.~\ref{fig:sigma:m0}, and
corresponds to the dominance of the $\tau^+ \tau^-$ decay channel.
The decrease of the cross sections for large $m_{1/2}$ is caused by
the reduction in the branching ratio for $\tilde e_L \to e^- \tilde \chi_2^0$
and is also a consequence of the smaller phase space available for selectron
pair production. In fact, with $m_0 = 80$ GeV, the production of
$\tilde e_L \tilde e_L$ pairs is only kinematically possible when
$m_{1/2} \leq 340$ GeV.

\begin{figure}[htb]
\begin{center}
\mbox{\epsfig{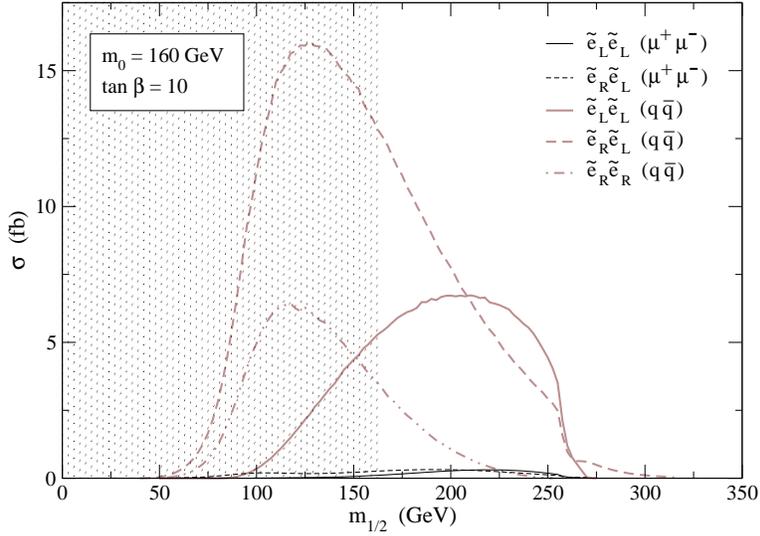}}
\caption{Cross sections (in fb) for 
$e^- e^- \to \tilde e \tilde e \to 
e^- \tilde \chi_1^0 e^- \tilde \chi_2^0
\to e^- \tilde \chi_1^0  e^- \tilde \chi_1^0 f \!\bar f$
(with unpolarised beams)
as a function of $m_{1/2}$, for $m_0 = 160$ GeV and $\tan \beta = 10$.}
\label{fig:sigma:m12-2}
\end{center}
\end{figure}

For scenario 2, we
fix $m_0 = 160$ GeV, $\tan \beta = 10$, $A_0=0$ (as in
Table~\ref{tab:in}), and plot the cross sections 
for different values of $m_{1/2}$ (see Fig.~\ref{fig:sigma:m12-2}).
The shaded area on the left of this figure
corresponds to values of $m_{1/2}$ for which $m_{\tilde \chi_1^0}$ is 
excluded~\cite{PDB}. The sudden decrease of the cross sections 
for $m_{1/2} \gtrsim 250$ GeV is caused by
both the dominance of the $\tau^+ \tau^-$ channel and the small phase space
available for selectron pair production. For $m_0 = 160$ GeV, the production of
$\tilde e_L \tilde e_L$ pairs is kinematically forbidden with $m_{1/2} \geq
270$ GeV. The production of $\tilde e_R \tilde e_R$ pairs is still allowed, but
for $m_{1/2} \geq 260$ GeV the $\tilde \chi_2^0$ is heavier than the $\tilde
e_R$ and the decay $\tilde e_R \to e^- \tilde \chi_2^0$ is not possible.

\begin{figure}[htb]
\begin{center}
\mbox{\epsfig{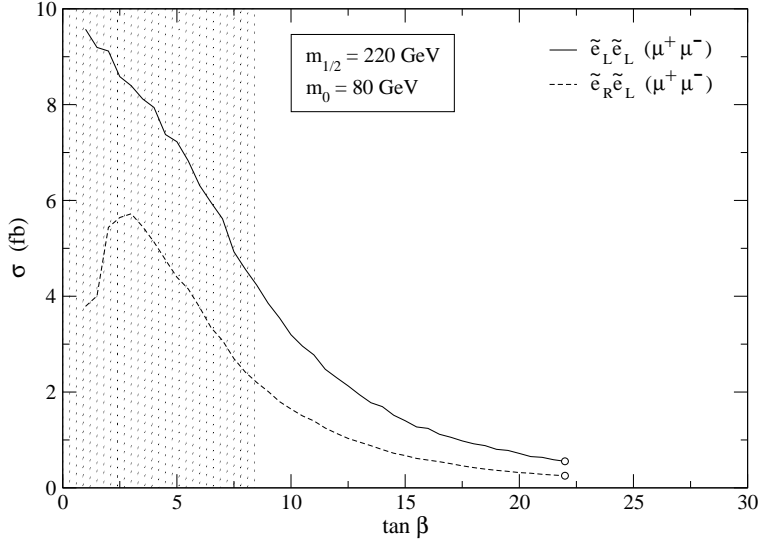}}
\caption{Cross sections (in fb) for 
$e^- e^- \to \tilde e \tilde e \to 
e^- \tilde \chi_1^0 e^- \tilde \chi_2^0
\to e^- \tilde \chi_1^0  e^- \tilde \chi_1^0 f \!\bar f$
(with unpolarised beams)
as a function of $\tan \beta$, for $m_{1/2}=220$ GeV and $m_0 = 80$ GeV.}
\label{fig:sigma:tanb}
\end{center}
\end{figure}

\begin{figure}[htb]
\begin{center}
\mbox{\epsfig{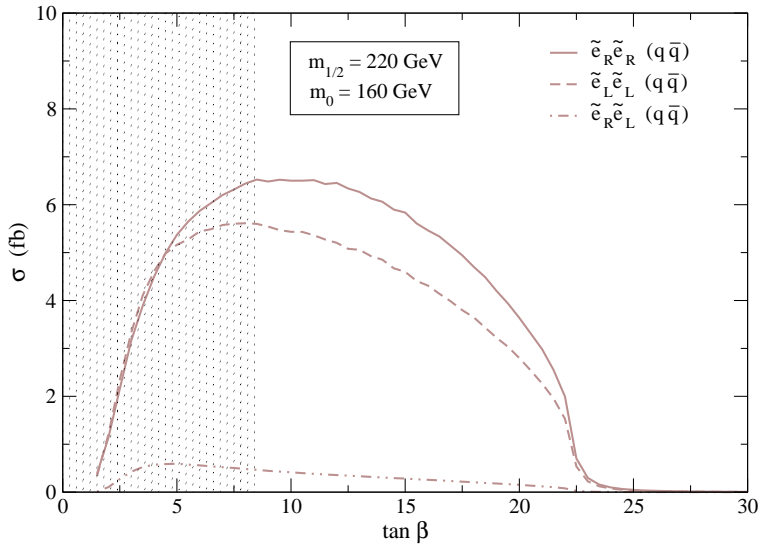}}
\caption{Cross sections (in fb) for 
$e^- e^- \to \tilde e \tilde e \to 
e^- \tilde \chi_1^0 e^- \tilde \chi_2^0
\to e^- \tilde \chi_1^0  e^- \tilde \chi_1^0 f \!\bar f$ 
(with unpolarised beams)
as a function of $\tan \beta$, for $m_{1/2}=220$ GeV and $m_0 = 160$ GeV.}
\label{fig:sigma:tanb-2}
\end{center}
\end{figure}

We also examine the dependence of the cross sections on $\tan \beta$ for
scenarios 1 and 2. Fixing $m_{1/2} = 220$ GeV, $m_0 = 80,160$ GeV, 
$A_0=0$ (as in Table~\ref{tab:in}), we plot in
Figs.~\ref{fig:sigma:tanb} and \ref{fig:sigma:tanb-2}, respectively, the cross
sections for different values of $\tan \beta$.
In both figures the shaded areas on the left are excluded by
the LEP bounds on the MSSM 
lightest Higgs mass\footnote{Since theoretical calculations of 
$m_{h^0}$ have an estimated error of 
2-3 GeV \cite{errhiggs}, we assume the conservative bound $m_{h^0} >111$ GeV.}, 
and in Fig.~\ref{fig:sigma:tanb}
the end points of the curves marked
with a circle correspond to values of $\tan \beta$
for which the LSP is no longer $\tilde \chi_1^0$. In both cases
the regions of parameter space for which the process
$e^- e^- \to \tilde e_L \tilde e_L \to e^- \tilde \chi_1^0 e^-
\tilde \chi_1^0 f \!\bar f$
is observable range up to $\tan \beta \simeq 22$.

The study of Figs.~\ref{fig:sigma:m0}--\ref{fig:sigma:tanb-2} reveals that
for most regions of mSUGRA parameter space where $\tilde e_L \tilde e_L$
production is kinematically accesible, either the $\mu^+ \mu^-$ or the $q \bar
q$ decay channels have observable cross sections. 
The only exceptions are narrow
intervals in $m_0$ and $m_{1/2}$, and also large $\tan \beta$ values, where the
$\tau^+ \tau^-$ channel completely dominates the
$\tilde \chi_2^0$ decays. On the other hand, the cross section for
$e^- e^- \to \tilde e_R \tilde e_R \to e^- \tilde \chi_1^0 e^- \tilde \chi_2^0
\to e^- \tilde \chi_1^0 e^- \tilde \chi_1^0 f \!\bar f$ is always small, as a
consequence of the small branching ratio for 
$\tilde e_R \to e^- \tilde \chi_2^0$ (in addition, this decay mode is not
possible when $m_{\tilde e_R} \lesssim m_{\tilde \chi_2^0}$).
Finally, we remark that, as shown in
Figs.~\ref{fig:sigma:m0}--\ref{fig:sigma:m12-2}, for larger values of $m_{1/2}$
and/or $m_0$ selectron pair production and, in general, the production of
supersymmetric particles, is not kinematically allowed at the assumed CM energy
of 500 GeV, and a higher energy collider (or the TESLA upgrade) would be
required for studies involving a heavier SUSY spectrum.

\section{Reconstruction of the selectron masses}
\label{sec:4}

To reconstruct the selectron masses we use as
input the 4-momenta of the detected particles (the two electrons and the
$f \! \bar f$ pair), the CM energy and the 
$\tilde \chi_1^0$ and $\tilde \chi_2^0$ masses, which we assume known from
other experiments \cite{TESLA,martyn}.
Remarkably, the values of both selectron masses are not needed anywhere. 
Let us first fix our notation by labelling the two electrons in the
final state as ``electron $a$'' and ``electron $b$'' (either choice is
equivalent) and their momenta as $p_a=(E_a,\vec p_a)$ and $p_b=(E_b,\vec p_b)$.
The selectrons to which they
correspond are then called ``selectron $a$'' and
``selectron $b$'', and the $\tilde \chi_1^0$ resulting from the decay of the
latter are also labelled as ``$a$'' and ``$b$'', with momenta
$q_a=(E_a',\vec q_a)$ and $q_b=(E_b',\vec q_b)$, respectively.
The momentum of the
$f \! \bar f$ pair is denoted as $p_{f \! \bar f}=(E_{f \! \bar f},
\vec p_{f \! \bar f})$.

In general, it is necessary to have as many kinematical relations as unknown
variables in order to determine the momenta of the undetected particles.
In our case, there are 8 unknowns (the 4 components of $q_a$ and
$q_b$) and 8 constraints. From energy and momentum conservation and
the fact that the two $\tilde \chi_1^0$ are on shell, we have
\begin{eqnarray}
E_a'+E_b' & = & E_\mathrm{CM} - (E_a+E_b+E_{f \! \bar f}) \;, \nonumber \\
\vec q_a + \vec q_b & = & \vec \pmiss \;, \nonumber \\
q_a^2 & = & m_{\tilde \chi_1^0}^2 \;, \nonumber \\
q_b^2 & = & m_{\tilde \chi_1^0}^2 \;,
\label{ec:con1}
\end{eqnarray}
with $\vec \pmiss$ the missing momentum. Another equation is obtained from the
decay of the intermediate $\tilde \chi_2^0$.
Since we do not know {\em a priori} which selectron decays to
$\tilde \chi_2^0$, we must consider
both possibilities. If we assume that it is the selectron ``$a$'', we have
\begin{equation}
(q_a+p_{f \! \bar f})^2 = m_{\tilde \chi_2^0}^2 \hspace*{4.7cm}
\mbox{[case (a)]} \;,
\label{ec:con2a}
\end{equation}
or, assuming it is the selectron ``$b$'',
\begin{equation}
(q_b+p_{f \! \bar f})^2 = m_{\tilde \chi_2^0}^2 \hspace*{4.7cm}
\mbox{[case (b)]} \;.
\label{ec:con2b}
\end{equation}
In order to set the last kinematical constraint, we make the
hypothesis that in the $e^- e^-$ scattering two particles of equal mass are
produced\footnote{This applies for $\tilde e_L \tilde e_L$ and
$\tilde e_R \tilde e_R$
production, but not for $\tilde e_R \tilde e_L$, in which case the selectron
masses cannot be reconstructed.}.
This implies that in one case we have
\begin{equation}
E_a+E_a'+E_{f \! \bar f} = E_b+E_b' \hspace*{3cm} \mbox{[case (a)]}
\label{ec:con3a}
\end{equation}
while in the other
\begin{equation}
E_a+E_a' = E_b+E_b'+E_{f \! \bar f} \hspace*{3cm} \mbox{[case (b)]} \;.
\label{ec:con3b}
\end{equation}

It is worthwhile remarking that ISR, beamstrahlung, 
particle width effects and detector
resolution degrade the determination of the $\tilde \chi_1^0$
momenta. The ISR and beamstrahlung
modify the effective beam energies, and therefore the first two of
Eqs.~(\ref{ec:con1}) do not hold exactly. The finite detector resolution
also implies that Eqs.~(\ref{ec:con1}--\ref{ec:con3b}) are only approximate.
Moreover, the off-shellness of the selectrons and of the $\tilde \chi_2^0$
have a non-negligible influence on Eqs.~(\ref{ec:con2a}--\ref{ec:con3b}).
As a result of all these effects, in some cases
the solutions of these 
equations are imaginary, even when the
$\tilde \chi_2^0$ is ``correctly'' assigned to a selectron.
In order to find a real solution in any circumstance, 
we force the discriminants of the second degree 
equations involved to be non-negative.

Each set of equations -- Eqs.~(\ref{ec:con1},\ref{ec:con2a},\ref{ec:con3a})
and Eqs.~(\ref{ec:con1},\ref{ec:con2b},\ref{ec:con3b}) --
yields two solutions for $q_a$ and $q_b$.
Among these four
solutions, we have to select one, which in turn gives a value for
the reconstructed mass. To do so, we observe the following criteria:
\begin{enumerate}
\item[(I)] We first eliminate the unphysical
solutions yielding negative reconstructed selectron masses
$m_{\tilde e}^\mathrm{rec}$.
\item[(II)] We then discard the solutions in which the reconstructed mass
of the $\tilde \chi_2^0$, $m_{\tilde \chi_2^0}^\mathrm{rec}$, is too different
from the real
value\footnote{Although by construction $m_{\tilde \chi_2^0}^\mathrm{rec}
= m_{\tilde \chi_2^0}$, forcing the system of equations to have a real solution
relaxes this equality in some cases (see the preceeding paragraph).}:
for the $\mu^+ \mu^-$ channel we require
$|m_{\tilde \chi_2^0}^\mathrm{rec} - m_{\tilde \chi_2^0}| \leq 5$ GeV, and for
the $q \bar q$ channel 
$|m_{\tilde \chi_2^0}^\mathrm{rec} - m_{\tilde \chi_2^0}|
\leq 10$ GeV.
\item[(III)] Among the remaining solutions, we take the one giving the
smallest value of $m_{\tilde e}^\mathrm{rec}$. This choice provides
the ``correct'' solution in most cases. 
If no solution is left after (II), the
event is discarded.
\end{enumerate}
The above procedure determines the momenta of the two unobserved
$\tilde \chi_1^0$,
identifying which selectron has decayed to $\tilde \chi_2^0$ as well,
allowing to reconstruct the selectron masses. The reconstruction also
allows to distinguish between the electrons resulting from 
$\tilde e \to e^- \tilde \chi_1^0$, and $\tilde e \to e^- \tilde \chi_2^0$.
Let us define $E_1$ and $E_2$ as the energies of the electrons produced
in the decays $\tilde e \to e^- \tilde \chi_1^0$, and
$\tilde e \to e^- \tilde \chi_2^0$, respectively.
In the selectron rest frame, the electron energy is fixed by the kinematics of
the 2-body decay, and since the selectrons are spinless
particles, the decay is isotropic. 
Then, for $\tilde e_L \tilde e_L$, $\tilde e_R \tilde e_R$
production the energy distribution of
the electrons in the CM frame is flat, with end points at
\cite{feng1993,martyn2}
\begin{eqnarray}
E_i^\mathrm{max} & = & \frac{\sqrt s}{4}
\left( 1- \frac{m_{\tilde \chi_i^0}^2}{m_{\tilde e_{L,R}}^2} \right) (1 + \beta)
\;, \nonumber \\
E_i^\mathrm{min} & = & \frac{\sqrt s}{4}
\left( 1- \frac{m_{\tilde \chi_i^0}^2}{m_{\tilde e_{L,R}}^2} \right) (1 - \beta)
\;,
\label{ec:end}
\end{eqnarray}
where $\beta = \sqrt{1-4 m_{\tilde e_{L,R}}^2/s}$. For mixed selectron
production
the electron energy spectra are flat as well, but the expressions of the end
points are more involved. The results for the reconstruction are illustrated
for scenarios 1 and 2.

\subsection{Scenario 1}

\begin{figure}[htb]
\begin{center}
\epsfig{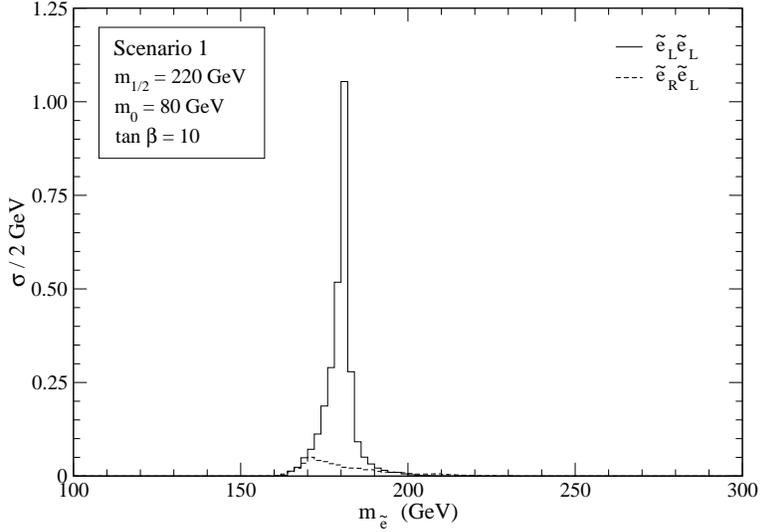}
\caption{Reconstructed selectron mass in scenario 1, for unpolarised beams.}
\label{fig:rec1}
\end{center}
\end{figure}

As seen from Fig.~\ref{fig:rec1}, in scenario 1
the reconstruction of the $\tilde e_L$ mass is quite effective, 
and a sharp peak at the true value
$m_{\tilde e_L} = 181$ GeV (as given in Table \ref{tab:out}) is obtained.
On the other hand, the
distribution of the  $\tilde e_R \tilde e_L$ events is rather flat.
We notice that the criteria
(I,II) in the reconstruction procedure operate as an efficient 
kinematical cut,
suppressing the unwanted $\tilde e_R \tilde e_L$ production, 
while for $\tilde e_L \tilde e_L$ the effect
is much smaller (cf. Table \ref{tab:rec1}).
Assuming an integrated luminosity of 100 fb$^{-1}$, the number of events
collected in the detector is sufficiently large to allow the observation of
a positive signal,
with a mass distribution concentrated at $m_{\tilde e_L} = 181$ GeV.
The use of negative beam polarisation $P_1 = P_2 = -0.8$ improves these
results,
increasing the $\tilde e_L \tilde e_L$ cross section by a factor of 3.24 and
reducing  the $\tilde e_R \tilde e_L$ cross section by a factor of
0.36 (see Table \ref{tab:rec1} and Appendix \ref{sec:B}).

\begin{table}[htb]
\begin{center}
\begin{tabular}{ccccccc}
& ~ & \multicolumn{2}{c}{$P_{00}$} & ~ & \multicolumn{2}{c}{$P_{--}$} \\
& & before & after & & before & after \\
\hline
$\tilde e_L \tilde e_L$ & & 3.25 & 2.88 & & 10.52 & 9.32 \\
$\tilde e_R \tilde e_L$ & & 1.66 & 0.47 & & 0.60 & 0.17
\end{tabular}
\caption{Cross sections (in fb) for selectron pair production in scenario 1.
We quote results before and after
reconstruction, for unpolarised beams ($P_{00}$) and for
$P_1 = P_2 = -0.8$ ($P_{--}$).}
\label{tab:rec1}
\end{center}
\end{table}

\begin{figure}[htb]
\begin{center}
\epsfig{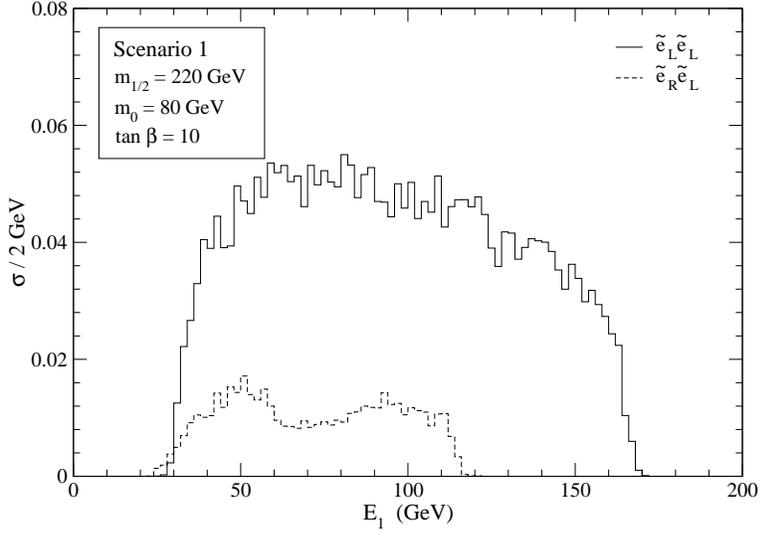}
\caption{Kinematical distribution of $E_1$ in scenario 1, for unpolarised
beams.}
\label{fig:rec1-E1}
\end{center}
\end{figure}

\begin{figure}[htb]
\begin{center}
\epsfig{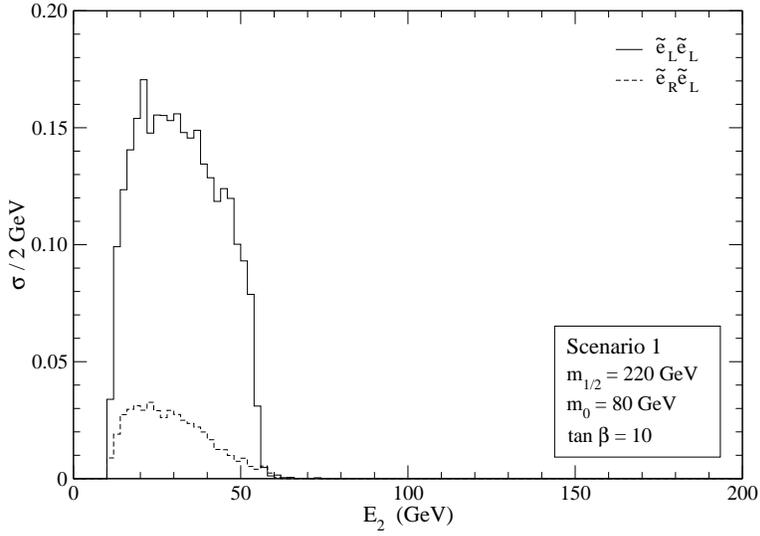}
\caption{Kinematical distribution of $E_2$ in scenario 1, for unpolarised
beams.}
\label{fig:rec1-E2}
\end{center}
\end{figure}

The particles observed in the kinematical distribution in Fig. \ref{fig:rec1}
are scalars, as can be easily deduced for instance from the angular
distributions for the production and the decay \cite{inprep}. The same can also
be concluded from the analysis of the electron energy spectra, which also
provides additional independent determinations of their mass.
From Eqs.~(\ref{ec:end}),
the expected end points of the distributions are $E_1^\mathrm{min} = 30$ GeV,
$E_1^\mathrm{max} = 166$ GeV and
$E_2^\mathrm{min} = 10$ GeV, $E_2^\mathrm{max} = 55$ GeV.
The kinematical distributions of $E_1$ and $E_2$ are shown
in Figs.~\ref{fig:rec1-E1} and \ref{fig:rec1-E2}, respectively.
We have included the $\tilde e_L \tilde e_L$ signal, as well as the
$\tilde e_R \tilde e_L$ background.
The energy spectrum of the electrons is smeared by ISR,
beamstrahlung, detector resolution and particle width effects.
However, the end points can be clearly observed in both plots,
further supporting the evidence for $\tilde e_L \tilde e_L$ production.

\subsection{Scenario 2}

\begin{figure}[htb]
\begin{center}
\epsfig{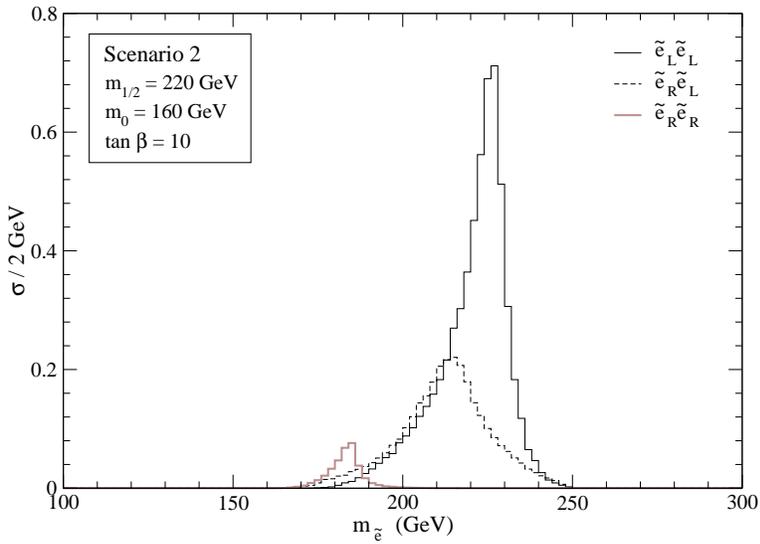}
\caption{Reconstructed selectron masses in scenario 2, for unpolarised beams.}
\label{fig:rec2}
\end{center}
\end{figure}

In this scenario, the $\tilde e_L \tilde e_L$ events produce a peak at
$m_{\tilde e_L} = 227$ GeV, and the $\tilde e_R \tilde e_R$ events yield
a tiny
peak around $m_{\tilde e_R} = 185$ GeV, as depicted in Fig.~\ref{fig:rec2}. 
Noticeably, the distribution 
of the $\tilde e_R \tilde e_L$ events concentrates around $215$
GeV. This behaviour is a result of the smaller ratio
$(m_{\tilde e_L}-m_{\tilde e_R})/(m_{\tilde e_L}+m_{\tilde e_R})$: in this
scenario, the hypothesis of two particles produced with equal mass, used for the
reconstruction, becomes more accurate. For the same reason, in this case this
background is less reduced by the criteria (I,II) than in scenario 1.
We observe that the $\tilde e_L \tilde e_L$ peak
is broader than
in scenario 1, due to the larger $\tilde e_L$ width and the less precise
measurement of the energy for jets. Additionally, the distribution is slightly
concentrated on smaller $m_{\tilde e_R}$ values, as a consequence of the
procedure used for the reconstruction.

\begin{table}[htb]
\begin{center}
\begin{tabular}{cccccccccc}
& ~ & \multicolumn{2}{c}{$P_{00}$} & ~ & \multicolumn{2}{c}{$P_{--}$} & ~ & 
\multicolumn{2}{c}{$P_{++}$} \\
& & before & after & & before & after & & before & after \\
\hline
$\tilde e_L \tilde e_L$ & & 6.52 & 5.71 & & 21.14 & 18.50 & & 0.26 & 0.23 \\
$\tilde e_R \tilde e_R$ & & 0.41 & 0.34 & & 0.02 & 0.01 & & 1.34 & 1.10 \\
$\tilde e_R \tilde e_L$ & & 5.41 & 2.63 & & 1.95 & 0.95 & & 1.95 & 0.95
\end{tabular}
\caption{Cross sections (in fb) for selectron pair production in scenario 2.
We quote results before and after reconstruction, for unpolarised beams
($P_{00}$), for $P_1 = P_2 = -0.8$ ($P_{--}$) and for
$P_1 = P_2 = 0.8$ ($P_{++}$).}
\label{tab:rec2}
\end{center}
\end{table}

The use of negative beam polarisation enhances the $\tilde e_L \tilde e_L$
signal and reduces the $\tilde e_R \tilde e_L$ background as in scenario 1,
practically eliminating $\tilde e_R \tilde e_R$ production (see
Table~\ref{tab:rec2}). Conversely, the use of positive beam
polarisation $P_1 = P_2 = 0.8$ enhances $\tilde e_R \tilde e_R$ by a factor of
3.24, suppressing
$\tilde e_L \tilde e_L$ and $\tilde e_R \tilde e_L$ 
by factors of 0.04 and 0.36,
respectively, and making the $\tilde e_R \tilde e_R$ peak clearly visible in
this case, as shown in Appendix \ref{sec:B}.

\begin{figure}[htb]
\begin{center}
\epsfig{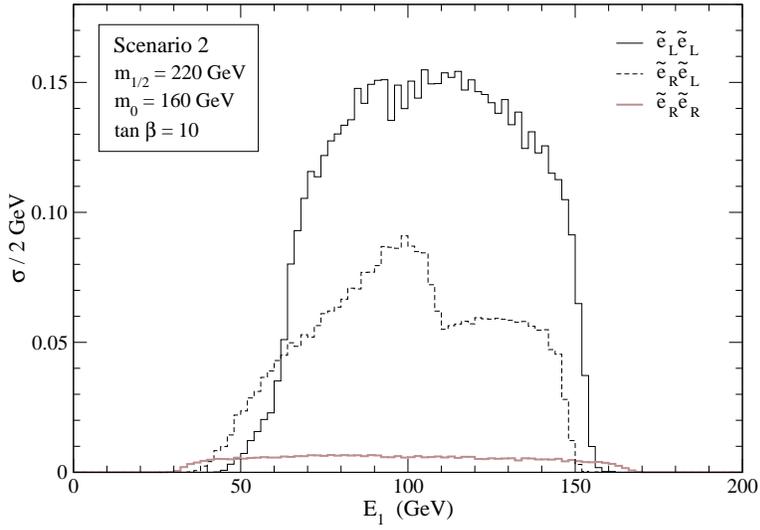}
\caption{Kinematical distribution of $E_1$ in scenario 2, for unpolarised
beams.
\label{fig:rec2-E1}}
\end{center}
\end{figure}

\begin{figure}[htb]
\begin{center}
\epsfig{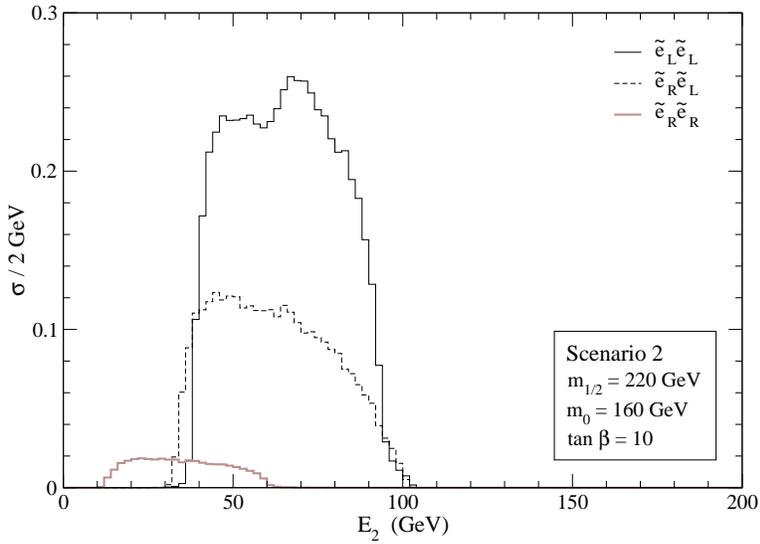}
\caption{Kinematical distribution of $E_2$ in scenario 2, for unpolarised
beams.
\label{fig:rec2-E2}}
\end{center}
\end{figure}

The examination of the energy spectra in Figs.~\ref{fig:rec2-E1} and
\ref{fig:rec2-E2} confirms the evidence of $\tilde e_L \tilde e_L$ production.
From Eqs.~(\ref{ec:end}) we obtain
for the decay $\tilde e_L \to e^- \tilde \chi_1^0$ the end points
$E_1^\mathrm{min} = 63$ GeV, $E_1^\mathrm{max} = 153$ GeV,
and for $\tilde e_L \to e^- \tilde \chi_2^0$ the limits
$E_2^\mathrm{min} = 38$~GeV, $E_2^\mathrm{max} = 93$ GeV.
In this scenario the $\tilde e_R \tilde e_L$ background is sizeable, and
negative beam polarisation might be needed in order to clearly determine
the end points of the electron energy distributions.
(Note that the large distortion in the energy spectra for $\tilde e_R \tilde
e_L$ is a result of the reconstruction process, which is devised for
the $\tilde e_L \tilde e_L$ and $\tilde e_R \tilde e_R$ signals.)
The electron energy distributions in $\tilde e_R \tilde e_R$ production 
display the end points at the expected energies
$E_1^\mathrm{min} = 32$ GeV, $E_1^\mathrm{max} = 166$ GeV,
$E_2^\mathrm{min} = 12$ GeV and $E_2^\mathrm{max} = 60$ GeV. For an integrated
luminosity of 100 fb$^{-1}$, these end points may be hard to observe even with
positive beam polarisation, because of the small cross section and the poor
statistics for this process. However, for larger luminosities the end points
could be observed more clearly (see Appendix \ref{sec:B}).

\section{Conclusions}\label{sec:5}

In this paper we have focused on the experimental test of the Majorana
nature of the neutralinos, studying selectron pair production in $e^- e^-$
scattering. These processes can only take place if, as it is predicted in most
SUSY models, the neutralinos are Majorana particles. Motivated by this issue,
we have demonstrated
that it is possible to reconstruct the selectron masses in the processes
$e^- e^- \to \tilde e_L \tilde e_L, \tilde e_R \tilde e_R$, using a relatively
rare channel with one selectron
decaying to $e^- \tilde \chi_1^0$ and the other one decaying via
$\tilde e \to e^- \tilde \chi_2^0 \to e^- \tilde \chi_1^0 f \!\bar f$.
The reconstruction can be done using as input
the 4-momenta of the detected particles, the CM energy and the 
$\tilde \chi_1^0$ and $\tilde \chi_2^0$ masses, which are assumed to be known.
In addition, the reconstruction procedure
allows for a determination of all final state momenta, which in turn has other
potential applications \cite{inprep}.
We have illustrated our results in two mSUGRA scenarios.
In the first scenario we have considered the channel $f \!\bar f = \mu^+ \mu^-$,
and in the second one the channel $f \!\bar f = q \bar q$. In the calculations
we have
taken into account ISR, beamstrahlung and particle width effects, and we have
performed a simple simulation of the energy resolution of the detector. We have
included the main background from $\tilde e_R \tilde e_L$ production as well.
We have shown that for the kinematically allowed processes the masses
can be successfully reconstructed in both
scenarios, yielding a clear peak in the case of $\tilde e_L \tilde e_L$
production and a smaller peak for $\tilde e_R \tilde e_R$ production.
The examination of the electron energy spectra shows that the produced
particles are scalars, making it obvious
that the observed signal corresponds to selectron pair production.
Additionally, the electron energy distributions display end points at the
energies expected for $\tilde e_L \tilde e_L$ and $\tilde e_R \tilde e_R$
production in each of the cases analysed, providing independent determinations
of the selectron masses and further supporting the evidence for
selectron pair production. Obviously, the hypothesis of selectron pair
production can also be confirmed by comparing the reconstructed masses with the
selectron masses precisely measured in other processes
\cite{feng:2001,blochinger:2002}.

In $e^- e^-$ collisions, selectron pair
production requires the exchange of a Majorana neutralino in the $t$ channel,
but does not prove that {\em all} the neutralinos are Majorana particles.
However, if the neutralino mixing matrix is precisely reconstructed in 
other experiments \cite{zerwas2,zerwas1,bartl2},
the measurement of the $\tilde e_L \tilde e_L$ and 
$\tilde e_R \tilde e_R$ cross
sections may allow us to identify the contributions of the different neutralino
mass eigenstates. In mSUGRA scenarios as those here considered,
where the two lightest neutralino mass eigenstates are
gaugino-like, $\tilde e_R \tilde e_R$ production is mainly mediated by the
exchange of $\tilde \chi_1^0$, with a smaller contribution from $\tilde
\chi_2^0$, the opposite occurring for $\tilde e_L \tilde e_L$ production.
Then, in these scenarios the measurement of one or both cross
sections can prove the exchange of Majorana neutralinos 
$\tilde \chi_1^0$ and $\tilde \chi_2^0$ (the analogous can be done in
more general scenarios where the reverse situation occurs, that is, when
$\tilde \chi_1^0$ is dominated by the wino component and  $\tilde
\chi_2^0$ is bino-like). The contribution to the cross sections
of the neutralinos with large higgsino-like components ($\tilde \chi_3^0$ and
$\tilde \chi_4^0$ in our case) is expected to be smaller than the uncertainty
in the measurement of the cross sections. 

It is worth noting here that
in addition to selectron pair production in $e^- e^-$ collisions, the processes
$e^+ e^- \to \tilde \chi_1^0 \tilde \chi_i^0 \to \tilde \chi_1^0 \tilde
\chi_1^0 l^+ l^-$, with $i=2,3,4$ and $l=e,\mu$,
may also shed light on the Dirac or Majorana nature of the neutralinos
\cite{petcov,bilenky,gudrid}.
The energy distributions of the final state charged leptons are sensitive to
the Dirac or Majorana nature of the decaying neutralino $\tilde \chi_i^0$,
allowing for the construction of lepton energy asymmetries, which vanish for a
Majorana $\tilde \chi_i^0$ but may have a nonzero value if $\tilde \chi_i^0$ is
a Dirac particle. Unfortunately,
these processes suffer from severe backgrounds from slepton pair production
$e^+ e^- \to \tilde l \tilde l \to l^+ l^- \tilde \chi_i^0 \tilde \chi_i^0$,
as well as chargino pair
$e^+ e^- \to \tilde \chi_i^+ \tilde \chi_j^- \to l^+ l^- \nu_l \bar \nu_l
\tilde \chi_i^0 \tilde \chi_i^0$ and $W^+ W^-$ production.
Moreover, for the case of
$\tilde \chi_1^0 \tilde \chi_3^0$
and $\tilde \chi_1^0 \tilde \chi_4^0$ the production cross sections and decay
branching ratios are in general small. If these difficulties
can be overcome, these processes may give further evidence of the Majorana
or Dirac nature of $\tilde \chi_2^0$, $\tilde \chi_3^0$ and $\tilde \chi_4^0$.

In this work we have taken as example the TESLA collider with a CM energy of
500 GeV, but the analysis presented can be applied to the proposed TESLA upgrade
to 800 GeV and other future colliders with larger CM energies, like CLIC.
In this case, with the larger energies and luminosities the processes
studied will be relevant in a larger region of parameter space, which is not
kinematically accesible at 500 GeV. Due to the $t$-channel nature of these
processes, their cross sections remain sizeable at TeV energies and hence
allow us to investigate the Majorana nature of the neutralinos in the case of a
heavier SUSY spectrum too.
Finally, let us remark the importance of the reconstruction of the selectron
masses on its own. This novel reconstruction technique can also be used, for
instance, in selectron and smuon pair production in $e^+ e^-$ collisions. It
provides a new and independent measurement of the masses of these sfermions,
and can be used as a tool for the analysis of the angular distributions in the
production and decay of these sfermions, and for the study of their spin
\cite{inprep}.
In addition, the mass reconstruction may be used in order to discriminate
between these and other SUSY signals.
For larger CM energies, the same method can be applied to squark pair
production, with or without the help of flavour tagging.

\vspace{1cm}
\noindent
{\Large \bf Acknowledgements}

\vspace{0.4cm} \noindent
We thank A. Bartl and S. Hesselbach for enlightening discussions and for a
critical reading of the manuscript. We are also indebted to the Vienna group
for their kind hospitality.
This work has been supported by the European Community's Human Potential
Programme under contract HTRN--CT--2000--00149 Physics at Colliders.
A.M.T. acknowledges the support by 
{\it Funda\c c\~ao para a Ci\^encia e Tecnologia} under
the grant SFRH/BPD/11509/2002.

\appendix 
\setcounter{equation}{0}
\renewcommand{\theequation}{\thesection.\arabic{equation}}

\section{Lagrangian}\label{sec:A}

\subsection{Mass matrices}

At low energies, the charged slepton mass term in the weak eigenstate basis is
\begin{equation}
\mathcal{L}_{\tilde E}^\text{mass} = -\frac{1}{2} {\tilde E}' \,^\dagger
M^2_{\tilde{E}} \tilde E' \;,
\end{equation}
with the most general charged slepton mass matrix
\begin{equation}\label{mass:slepton:gen}
M^2_{\tilde{E}}\;=\;
\left(\begin{array}{cc}
M^2_{\tilde{E}_{LL}}&M^2_{\tilde{E}_{LR}}\\
M^2_{\tilde{E}_{RL}} & M^2_{\tilde{E}_{RR}}
\end{array}\right).
\end{equation}
Omitting the flavour dependence, the above submatrices are given by~\cite{romao}
\begin{eqnarray}
M^2_{\tilde{E}_{LL}} & = &
  \frac{1}{2}\;v_1^2\; Y_e^* Y_e^T + M_{\tilde{L}}^2 -
  \frac{1}{2}\;m_Z^2 \, \cos  2\beta \;(1-2\sin^2 \theta_W)
  \, \unit\;, \nonumber \\
M^2_{\tilde{E}_{RR}} & = &
  \frac{1}{2}\;v_1^2\;Y_e^T Y_e^* + M_{\tilde{E}_R}^2-
  m_Z^2 \, \cos  2 \beta \; \sin^2 \theta_W \, \unit \;, \nonumber \\
M^2_{\tilde{E}_{LR}} & = &
  {M^{2 \dagger}_{\tilde{E}_{RL}}} = 
  \frac{v_1}{\sqrt{2}}\;(Y^A_e)^* - \mu \;\frac{v_2}{\sqrt{2}}\;Y_e^*\;,
\label{slepton:mass:2ELR}
\end{eqnarray}
where $Y_e$ are the charged lepton Yukawa couplings, $M_{\tilde{L}}$
and $M_{\tilde{E}_R}$ the soft breaking scalar masses for the left and
right handed sleptons, respectively, $Y^A_e$ the soft trilinear
terms and $\unit$ is the $3 \times 3$ identity matrix in flavour space.
The trilinear couplings can be decomposed as
$(Y^A)_{ij} \equiv A_{ij} \; Y_{ij}$, with no summation over $i,j$.
The slepton mass matrix is diagonalised by a $6 \times 6$ rotation, 
\begin{equation}\label{slepton:rotation}
(M_{\tilde{E}}^\text{diag})^2 \;=\; R_{\tilde{E}} \;M_{\tilde{E}}^2\;
R_{\tilde{E}} \,^{\dagger} \;,
\end{equation}
with $\tilde E = R_{\tilde E} \, \tilde E'$ the mass eigenstates.
Neglecting the mixing between generations, the matrices in 
Eqs.~(\ref{mass:slepton:gen},\ref{slepton:mass:2ELR})
are diagonal in flavour space, and only $LR$ mixing is present.
Hence, the former relation between mass and weak 
interaction eigenstates
can be simply expressed for each flavour as
\begin{equation}\label{mass:slepton:Rl}
\tilde l = R^{\tilde l} \; \tilde l' \;,
\end{equation}
where $R^{\tilde l}$ now denotes a $2 \times 2$ matrix, 
with $l=e,\, \mu,\, \tau$.
Using the conventions of Ref. \cite{spheno}, and assuming that the matrices in
Eqs.~(\ref{slepton:mass:2ELR}) are real,
\begin{equation}\label{mass:slepton:theta}
\left(
\begin{array}{c}
\tilde l_1 \\
\tilde l_2
\end{array} \right) = \left(
\begin{array}{cc}
\sin\theta_{\tilde l} & -\cos \theta_{\tilde l} \\
\cos \theta_{\tilde l} & \sin \theta_{\tilde l} 
\end{array} \right) \; \left(
\begin{array}{c}
\tilde l_L \\
\tilde l_R
\end{array} \right)\;,
\end{equation}
with $m_{\tilde l_1} < m_{\tilde l_2}$.

Squark-mediated interactions have a sub-dominant role in the parameter space
a\-na\-ly\-sed. Similarly to the slepton case, the 
$6 \times 6$ squark mass matrices can be diagonalised by a
unitary matrix $R_{\tilde Q}$,
\begin{equation}\label{squark:rotation}
(M_{\tilde Q}^{\text {diag}})^2 \;=\; R_{\tilde Q} \;M_{\tilde Q}^2\;
{R_{\tilde Q}}^{\dagger}\;,\quad  \tilde Q= \tilde U, \,\tilde D
\end{equation}
so that the relation between the weak (primed) and
mass eigenstates is given by
\begin{equation}\label{squark:rotation:specific}
\tilde{q}\;=\; R_{\tilde Q} \;\tilde{q}^\prime\;.
\end{equation}

The four Majorana neutralinos $\tilde \chi^0_i$ are mixtures of 
weak interaction eigenstates 
(bino, neutral wino and neutral higgsinos). In the 
basis where 
$(\psi^{0})^T=(\tilde{B},\tilde{W}^0_3,\tilde{H}_1^0,\tilde{H}_2^0)^T$,
the mass term is
\begin{equation}\label{neutr:mass:gen}
\mathcal{L}_{\tilde \chi^0}^\text{mass} \;=\;
-\frac{1}{2} \;(\psi^{0})^T\;M_{\tilde \chi^0} \; \psi^0 + \text{H. c.} \;,
\end{equation}
and the neutralino mass matrix can be written as
\begin{equation} \label{neut:mass:matrix}
M_{\tilde \chi^0} \; = \; \left(
\mbox{\footnotesize $\begin{array}{cccc}
m_1 & 0 & -m_Z\,\sin \theta_W \, \cos \beta &
 m_Z\,\sin \theta_W \, \sin \beta \\
0 & m_2 & m_Z\,\cos \theta_W \, \cos \beta &
 -m_Z\,\cos \theta_W \, \sin \beta \\
-m_Z\,\sin \theta_W \, \cos \beta &
 m_Z\,\cos \theta_W \, \cos \beta & 0 & -\mu \\
m_Z\,\sin \theta_W \, \sin \beta &
 -m_Z\,\cos \theta_W \, \sin \beta & -\mu & 0
\end{array} $}
\right)\;,
\end{equation}
where $m_{1,2}$ are the soft gaugino masses. This matrix
is diagonalised by
\begin{equation}\label{neut:rotation}
N^*\;M_{\tilde \chi^0}\;N^{-1}\;=\;M_{\tilde \chi^0}^{\text diag}\;.
\end{equation}

\subsection{Interaction terms}
In this section, we list the
interactions~\cite{susy:2,romao,gunion,hunter} relevant for the
calculation of the production and decay cross sections.
The fermion-sfermion-neutralino terms are given by
\begin{equation} 
\mathcal{L}_{\tilde{f_i} f \tilde{\chi}_j^0} =
\tilde{f_i}^* \;
\bar{\tilde \chi}_j^0 \left[ (C^{f}_L)_{ij} P_L + (C^{f}_R)_{ij} P_R
\right] f + 
\tilde{f_i} \;
\bar f \left[ (C^{f}_R)_{ij}^* P_L + (C^{f}_L)_{ij}^* P_R \right] 
\tilde \chi_j^0   \;.
\end{equation}
For the case of charged (s)leptons, the $C$ couplings are
\begin{eqnarray}
(C^{l}_L)_{ij} & = &
\frac{g}{\sqrt{2}} (N^*_{j2} + \tan \theta_W N^*_{j1}) \,
R^{\tilde l}_{i1} - Y_{e} \, N^*_{j3} \, R^{\tilde l}_{i2} \;, \nonumber \\
(C^{l}_R)_{ij} & = &
-g \sqrt{2} \tan \theta_W N_{j1} \, R^{\tilde l}_{i2} -
Y_{e} \, N_{j3} \, R^{\tilde l}_{i1} \;, 
\end{eqnarray}
with $l=e,\, \mu,\, \tau$.
In our computations we have considered the limit where 
$Y_e=\text{diag}\,(0,0,h_\tau)$, so that the second term on the
r.h.s.~of the above equations is only present for the case of the
taus.

For the calculation of the decay $\tilde \chi_2^0 \to \tilde \chi_1^0 q \bar 
q$ we include squark-mediated interactions, though they have a minor 
importance. In this case, the couplings have a slightly more
cumbersome expression, which involves both quark and squark rotation
matrices. However, in this work, the effects of quark and squark
flavour mixing are not relevant, and one can safely ignore them. 
Moreover, as in the lepton sector, the Yukawa couplings of the two first
generations can be neglected.
In this limit, where $Y_u=\text{diag}\,(0,0,h_t)$,
$Y_d=\text{diag}\,(0,0,h_b)$ and $V_{CKM} \simeq \unit$, 
squark mixing is purely $LR$, and can be parametrised for every family
as in Eqs.~(\ref{mass:slepton:Rl}, \ref{mass:slepton:theta}).
With $u=u,\,c,\,t$ and $d=d,\, s,\,b$, 
the $C$ couplings read
\begin{eqnarray}
(C^{u}_L)_{ij} & = &
- \frac{g}{\sqrt{2}} (N^*_{j2} + \frac{1}{3} \tan \theta_W N^*_{j1}) \,
R^{\tilde u}_{i1} - Y_u \, N^*_{j4} \, R^{\tilde u}_{i2}\;, \nonumber \\
(C^{u}_R)_{ij} & = &
g \sqrt{2} \; \frac{2}{3} \tan \theta_W N_{j1} \, R^{\tilde u}_{i2} 
- Y_u \, N_{j4} \, R^{\tilde u}_{i1}\;,\nonumber \\
(C^{d}_L)_{ij} & = &
\frac{g}{\sqrt{2}} (N^*_{j2} - \frac{1}{3} \tan \theta_W N^*_{j1}) \,
R^{\tilde d}_{i1} - Y_{d} \, N^*_{j3} \, R^{\tilde d}_{i2} \;, \nonumber \\
(C^{d}_R)_{ij} & = &
-g \sqrt{2} \; \frac{1}{3} \tan \theta_W N_{j1} \, R^{\tilde d}_{i2} - 
Y_{d} \, N_{j3} \, R^{\tilde d}_{i1}\;.
\end{eqnarray}

The $Z$-neutralino-neutralino interactions are parametrised by
\begin{equation}
\mathcal{L}_{Z \tilde\chi_i^0 \tilde \chi_j^0} =
\frac{g}{2 \cos \theta_W}  \, Z_\mu \, \left[ 
\bar{\tilde \chi}_i^0 \gamma^\mu \left( D_L^{ij} P_L + 
D_R^{ij} P_R \right) \tilde \chi^0_j \right]  \;,
\end{equation}
where
\begin{eqnarray}
D_L^{ij} & = & \frac{1}{2} 
\left(N_{i4} N^*_{j4} - N_{i3} N^*_{j3} \right)\;, \nonumber \\
D_R^{ij} & = & -(D_L^{ij})^* \;.
\end{eqnarray}

Higgs-mediated interactions play a marginal role in the processes 
here analysed,
and are only non-negligible for final states with $f \! \bar f = b \bar b,
\tau^+ \tau^-$.
Assuming no CP violation in the Higgs sector (so that there is no
mixture between the CP-even and the CP-odd states), 
the interaction of the neutralinos with the neutral CP-even physical 
Higgs can be written as
\begin{eqnarray}
\mathcal{L}_{H^0 \tilde\chi_i^0 \tilde \chi_j^0 } & = &
-\frac{g}{\sqrt{2}} \, \left({H^0} \cos \alpha - {h^0} \sin \alpha \right) \,
\bar{\tilde \chi}^0_i
\left[ Q_{ij} P_L + Q^*_{ij} P_R \right] \tilde \chi^0_j \nonumber\\
& & + \frac{g}{\sqrt{2}} \, \left({H^0} \sin \alpha + {h^0} \cos \alpha \right) 
\, \bar{\tilde \chi}^0_i
\left[ S_{ij} P_L + S^*_{ij} P_R \right] \tilde \chi^0_j\;,
\label{higgs:neutralino:int}
\end{eqnarray}
where $\alpha$ is the CP-even Higgs mixing angle, and as usual
$m_{h^0} < m_{H^0}$. In Eq.~(\ref{higgs:neutralino:int}) we have the couplings
\begin{eqnarray}\label{Higgs:neutralino:QS}
Q_{ij} &=& \frac{1}{2}\left(
N^*_{i2} -\tan \theta_W N^*_{i1} \right) N^*_{j3} + (i
\leftrightarrow j) \;,\nonumber \\
S_{ij} &=& \frac{1}{2}\left(
N^*_{i2} -\tan \theta_W N^*_{i1}\right) N^*_{j4} + (i
\leftrightarrow j)\;.
\end{eqnarray}
The interaction of the pseudoscalar with the neutralinos can be written as
\begin{equation}
\mathcal{L}_{A^0 \tilde\chi_i^0 \tilde \chi_j^0 } =
-\frac{i\,g}{\sqrt{2}} \, A^0 \;
\bar{\tilde \chi}^0_i
\left[ \, \left( Q_{ij} \sin \beta -S_{ij} \cos \beta \right) P_L
- \left( Q_{ij}^* \sin \beta -S_{ij}^* \cos \beta \right) P_R \,
\right] \tilde \chi^0_j \;,
\end{equation}
with the couplings as in Eq.~(\ref{Higgs:neutralino:QS}).
The Higgs-fermion-fermion Lagrangian can be expressed,
omitting the flavour dependence, as
\begin{eqnarray}
\mathcal{L}_{H \! f\! f} &=& 
- Y_u^{\text{diag}}
\left[ \left(H^0 \sin \alpha + h^0 \cos \alpha \right) \, \bar u u 
- i \, A^0 \cos \beta \; \bar u \gamma_5 u \right] \nonumber \\
& & - Y_d^{\text{diag}}
\left[ \left(H^0 \cos \alpha - h^0 \sin \alpha \right) \, \bar d d 
- i \, A^0 \sin \beta \; \bar d \gamma_5 d \right] \nonumber \\
& & - Y_e^{\text{diag}}
\left[ \left(H^0 \cos \alpha - h^0 \sin \alpha \right) \, \bar l l 
- i \, A^0 \sin \beta \; \bar l \gamma_5 l \right] \;.
\end{eqnarray}
The $H \! f\! f$ interactions in the above equation are proportional to the
fermion Yukawa couplings, and hence are suppressed
in all cases, the exception being the top and bottom quarks and the
$\tau$. Finally,
the $Z \! f \! f$ couplings are the standard ones,
\begin{equation}
\mathcal{L}_{Z \!f \! f} = 
-\frac{g}{\cos \theta_W} \, Z_\mu \; \bar f \gamma^\mu 
\left[ \left( I_3^f - Q^f \sin^2 \theta_W \right) P_L 
-Q^f \sin^2 \theta_W P_R \right] f  \;, 
\end{equation}
where $I_3^f$ and $Q^f$ are, respectively, the weak isospin and charge of
fermion $f$.

\newpage
\setcounter{equation}{0}
\section{Effect of beam polarisation}
\label{sec:B}

The TESLA design offers the possibility of electron polarisation up to
$\pm 80$\%,
which provides several advantages for our study, reducing the
$\tilde e_R \tilde e_L$ background and enhancing the $\tilde e_L \tilde e_L$ or
$\tilde e_R \tilde e_R$ signals.
The cross sections for selectron pair production with longitudinally polarised
beams can be related to the unpolarised cross sections in a very simple way.
Since selectron mixing is negligible,
for initial electrons with definite helicity the only
non-vanishing amplitudes for selectron pair production are
$e_L^- e_L^- \to \tilde e_L \tilde e_L$,
$e_R^- e_R^- \to \tilde e_R \tilde e_R$ and
$e_R^- e_L^- \to \tilde e_R \tilde e_L$.
This allows to write the cross sections for arbitrary
polarisations $P_1$, $P_2$ as
\begin{eqnarray}
\sigma_{\tilde e_L \tilde e_L} (P_1,P_2) & = & (1-P_1) (1-P_2) \;
\sigma_{\tilde e_L \tilde e_L} (0,0) \;, \nonumber \\
\sigma_{\tilde e_R \tilde e_R} (P_1,P_2) & = & (1+P_1) (1+P_2) \;
\sigma_{\tilde e_R \tilde e_R} (0,0) \;, \nonumber \\
\sigma_{\tilde e_R \tilde e_L} (P_1,P_2) & = & (1-P_1 P_2) \;
\sigma_{\tilde e_R \tilde e_L} (0,0) \;.
\label{ec:pol}
\end{eqnarray}
For the sake of simplicity, throughout this article we have plotted cross
sections for unpolarised beams. For the case of polarised electrons, the cross
sections are straightforward to obtain, using Eqs.~(\ref{ec:pol}).
However, it is very illustrative to plot some of the distributions presented
in Section \ref{sec:4} for the case of polarised beams, making apparent the
enhancement of the signals and the reduction of $\tilde e_R \tilde e_L$
production. We only consider the cases $P_1 = P_2 = -0.8$
and $P_1 = P_2 = 0.8$. The use of a polarisation
$P_1 = - P_2 = 0.8$ does not offer any advantage for our study.

\subsection{Negative beam polarisation}

\begin{figure}[!htb]
\begin{center}
\epsfig{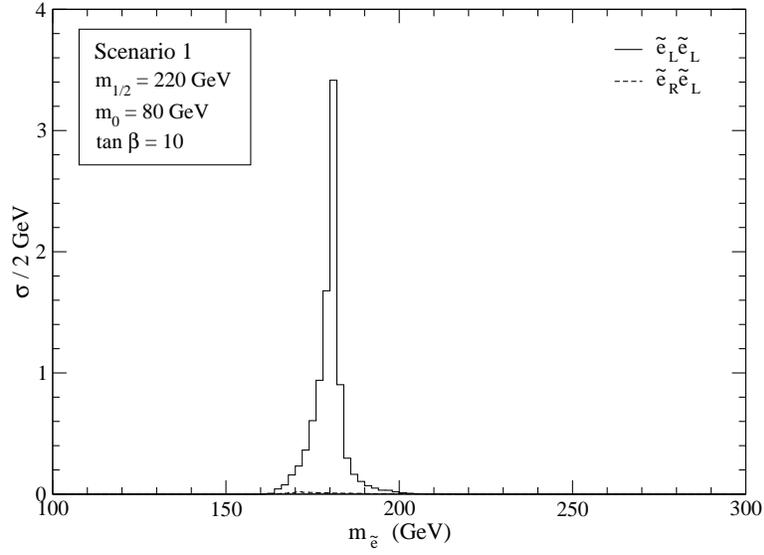}
\caption{Reconstructed selectron mass in scenario 1, for $P_1 = P_2 = -0.8$.}
\label{fig:rec1-P--}
\end{center}
\end{figure}

\begin{figure}[!htb]
\begin{center}
\epsfig{file=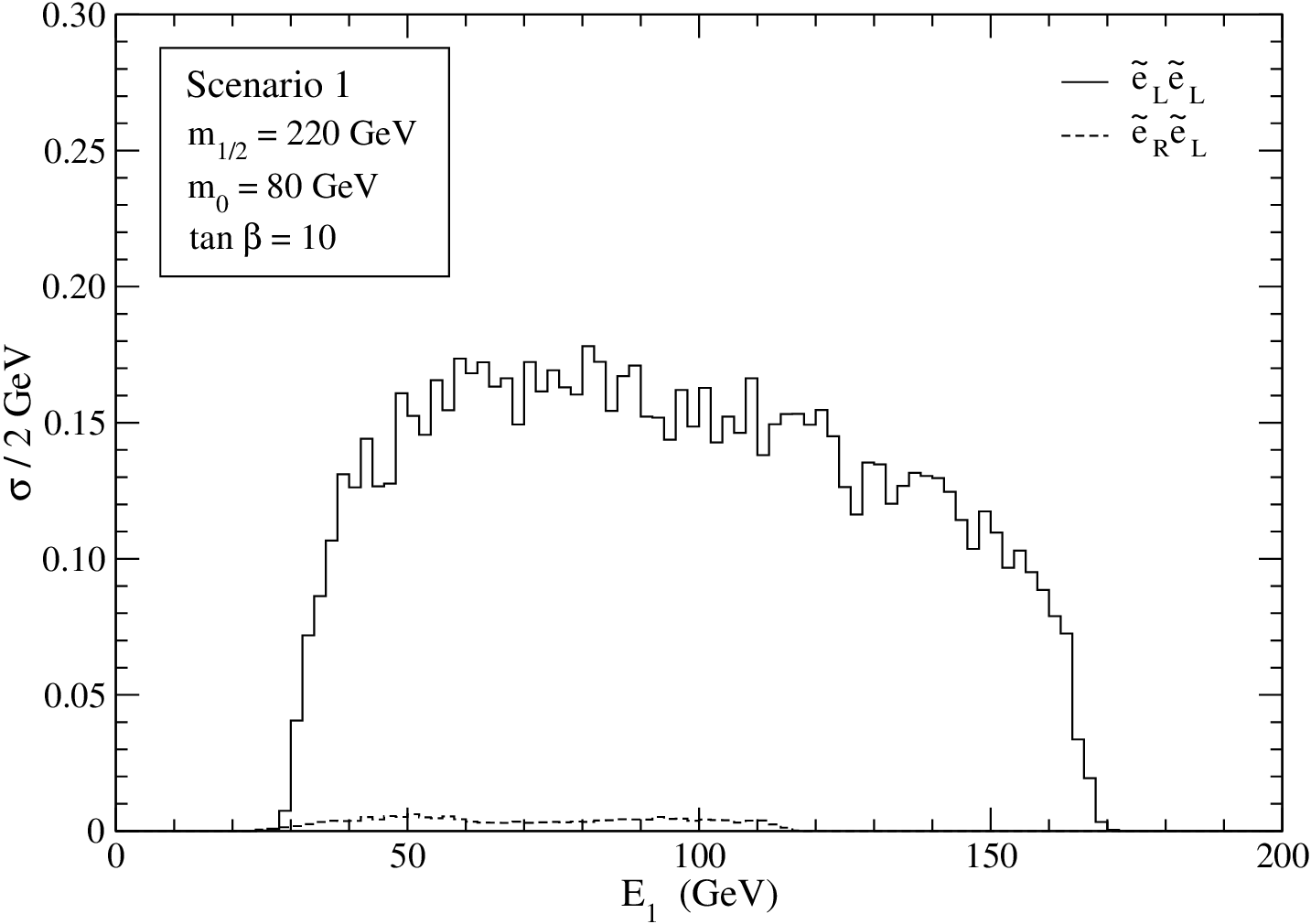,width=10cm,clip=}
\caption{Kinematical distribution of $E_1$ in scenario 1, for
$P_1 = P_2 = -0.8$.}
\label{fig:rec1-E1-P--}
\end{center}
\end{figure}

\begin{figure}[!htb]
\begin{center}
\epsfig{file=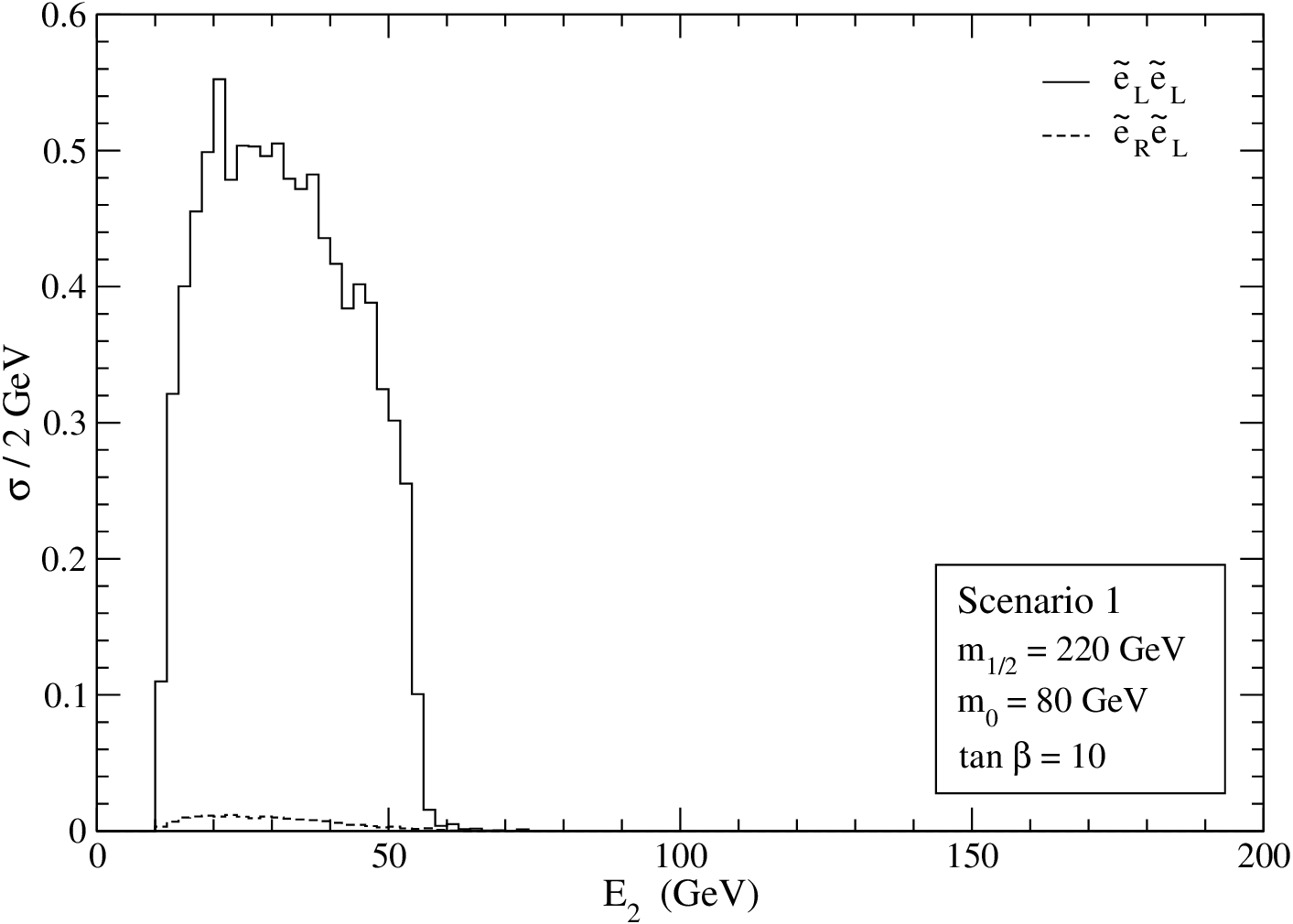,width=10cm,clip=}
\caption{Kinematical distribution of $E_2$ in scenario 1, for
$P_1 = P_2 = -0.8$.}
\label{fig:rec1-E2-P--}
\end{center}
\end{figure}

\begin{figure}[!htb]
\begin{center}
\epsfig{file=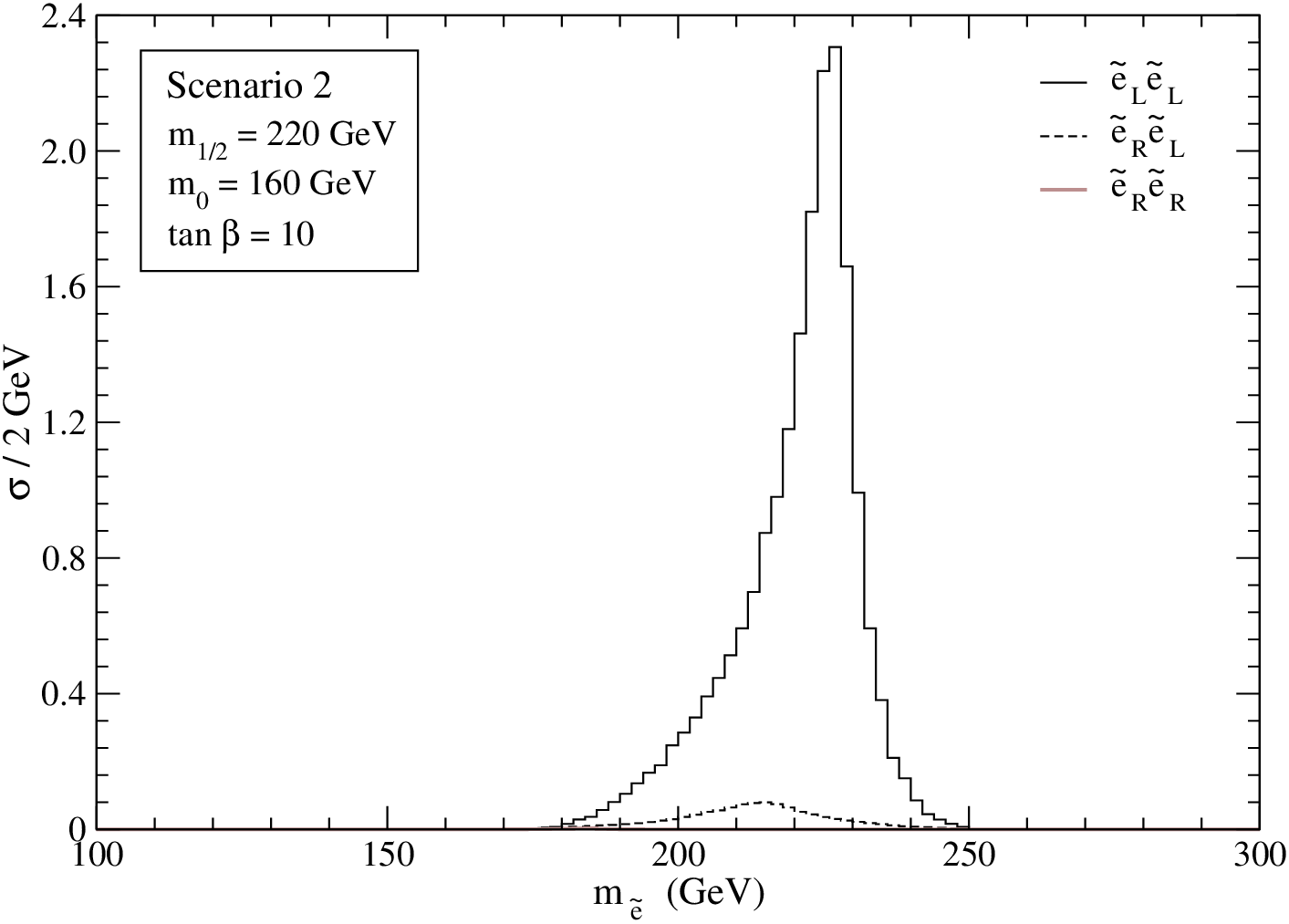,width=10cm,clip=}
\caption{Reconstructed selectron masses in scenario 2, for $P_1 = P_2 = -0.8$.}
\label{fig:rec2-P--}
\end{center}
\end{figure}

\begin{figure}[!htb]
\begin{center}
\epsfig{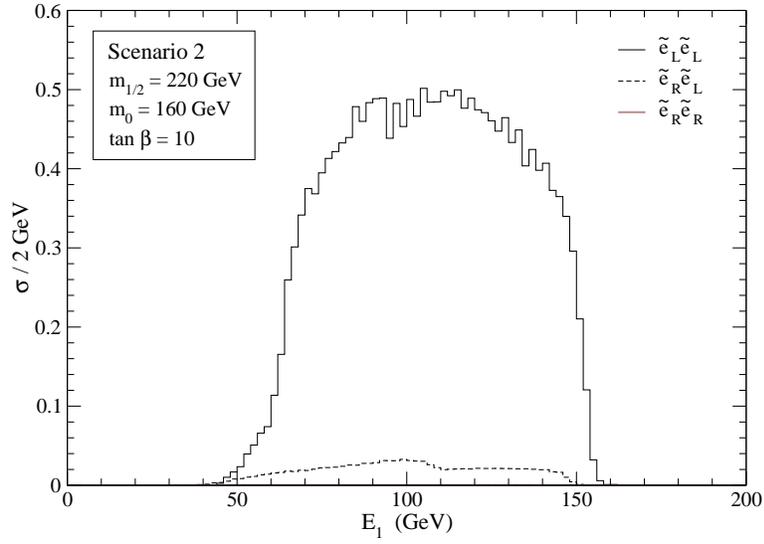}
\caption{Kinematical distribution of $E_1$ in scenario 2, for
$P_1 = P_2 = -0.8$.
\label{fig:rec2-E1-P--}}
\end{center}
\end{figure}

\begin{figure}[!htb]
\begin{center}
\epsfig{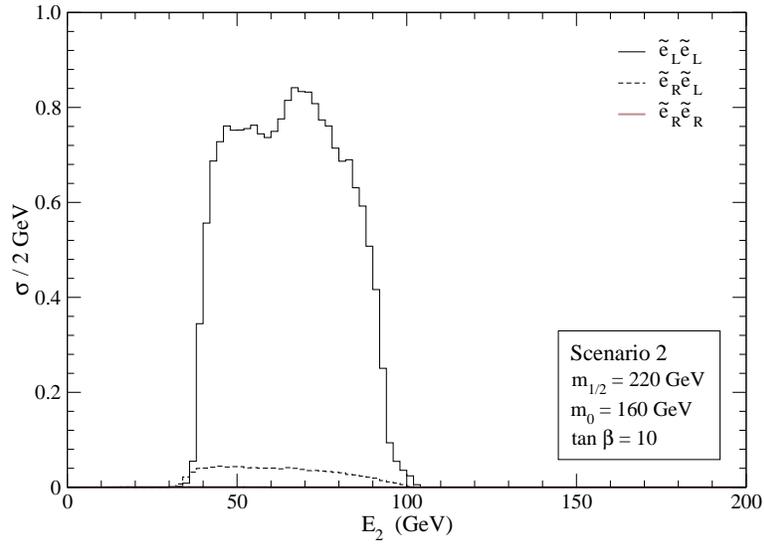}
\caption{Kinematical distribution of $E_2$ in scenario 2, for
$P_1 = P_2 = -0.8$.
\label{fig:rec2-E2-P--}}
\end{center}
\end{figure}

Negative beam polarisation $P_1 = P_2 = -0.8$ enhances the
$\tilde e_L \tilde e_L$ signal by a factor of 3.24, reducing the
$\tilde e_R \tilde e_R$ signal by a factor of 0.04 and the $\tilde e_R \tilde
e_L$ background by 0.36. The effect on the mass distributions and electron
energy spectra for scenario 1 can be seen in
Figs.~\ref{fig:rec1-P--}--\ref{fig:rec1-E2-P--}. For scenario
2, the corresponding cross sections are depicted in
Figs.~\ref{fig:rec2-P--}--\ref{fig:rec2-E2-P--}.

\clearpage
\subsection{Positive beam polarisation}

\begin{figure}[!htb]
\begin{center}
\epsfig{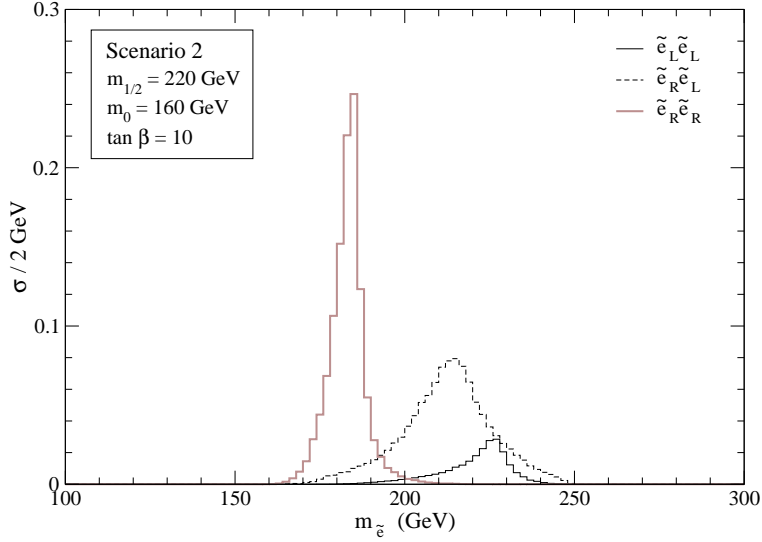}
\caption{Reconstructed selectron masses in scenario 2, for $P_1 = P_2 = 0.8$.}
\label{fig:rec2-P++}
\end{center}
\end{figure}

\begin{figure}[!htb]
\begin{center}
\epsfig{file=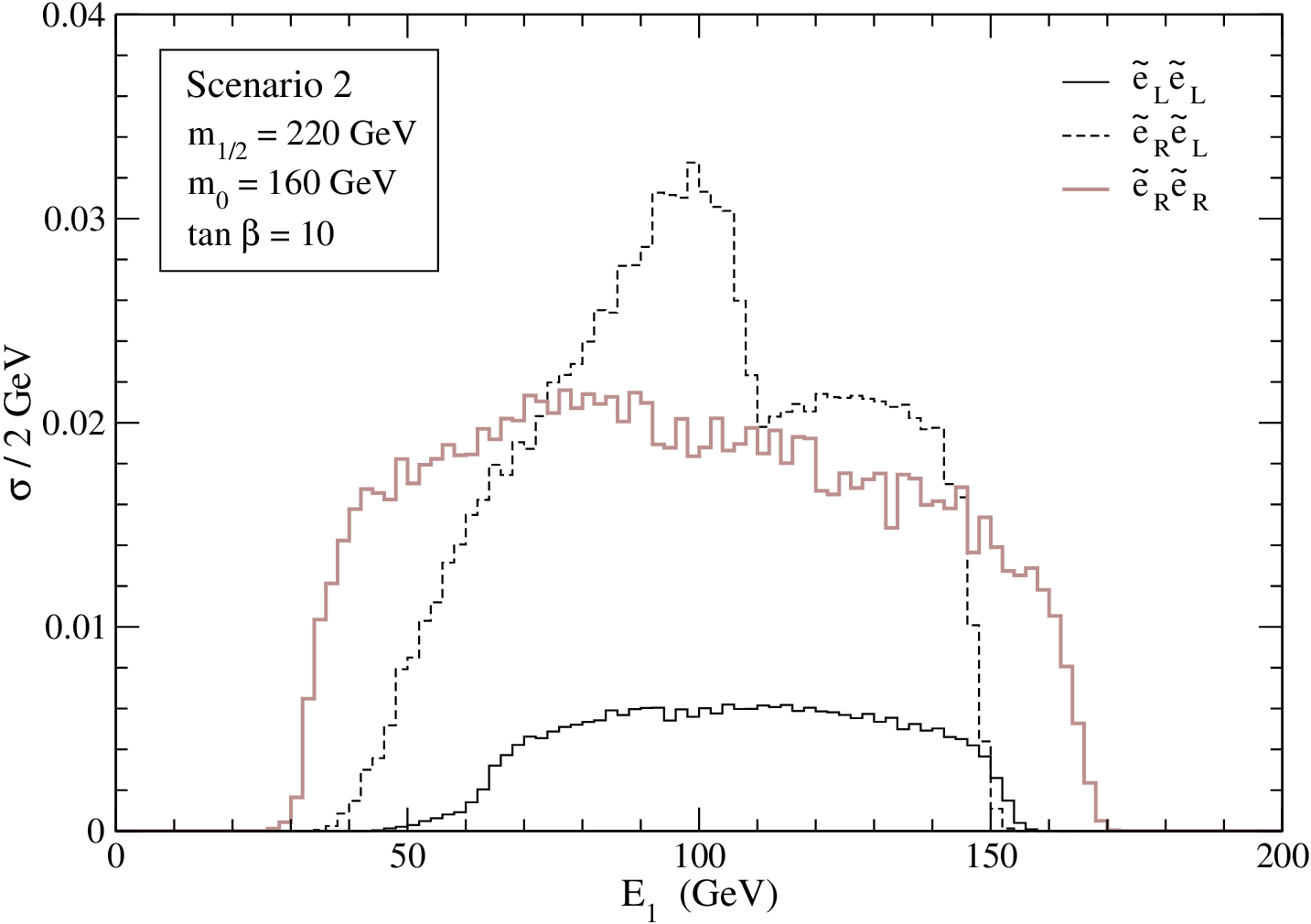,width=10cm,clip=}
\caption{Kinematical distribution of $E_1$ in scenario 2, for
$P_1 = P_2 = 0.8$.
\label{fig:rec2-E1-P++}}
\end{center}
\end{figure}

\begin{figure}[!htb]
\begin{center}
\epsfig{file=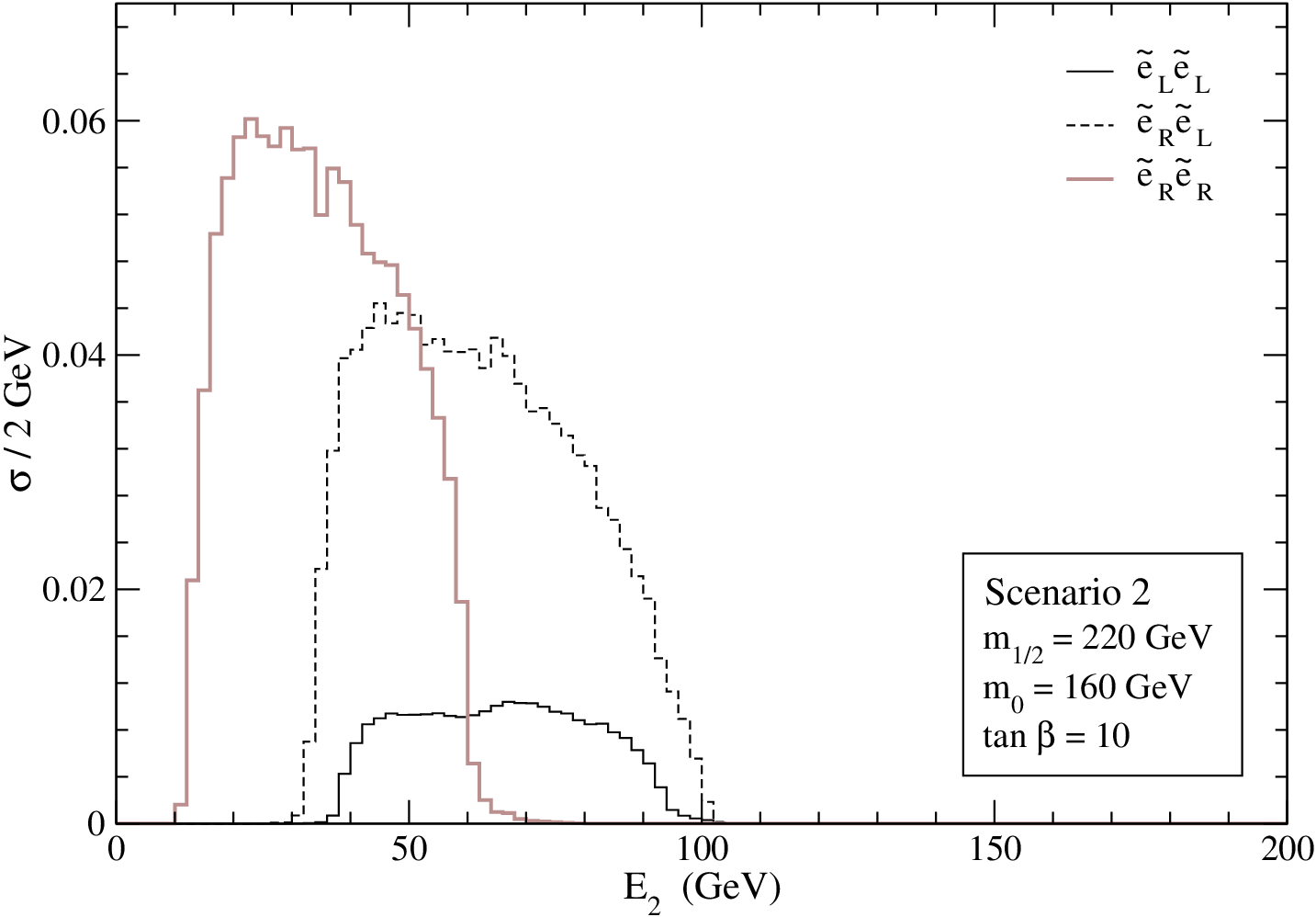,width=10cm,clip=}
\caption{Kinematical distribution of $E_2$ in scenario 2, for
$P_1 = P_2 = 0.8$.
\label{fig:rec2-E2-P++}}
\end{center}
\end{figure}

Positive beam polarisation $P_1 = P_2 = 0.8$ enhances the
$\tilde e_R \tilde e_R$ signal by a factor of 3.24, reducing the
$\tilde e_L \tilde e_L$ signal by a factor of 0.04 and the $\tilde e_R \tilde
e_L$ background by 0.36. In scenario 1, positive beam polarisation does not
offer any advantage, and we only present the plots for scenario 2,
in Figs.~\ref{fig:rec2-P++}--\ref{fig:rec2-E2-P++}.

\end{document}